%% file: main.tex
\documentclass[journal]{IEEEtran}
\usepackage{placeins}
\usepackage[utf8]{inputenc}
\usepackage{bm}
\usepackage{cite}
\usepackage[pdftex]{graphicx}
\usepackage{multirow}
\usepackage{subfigure}
\usepackage{caption}
\usepackage{multicol}
\usepackage{comment}
\usepackage[version=4]{mhchem}
\usepackage{color}
\newsavebox{\bigimage}
\usepackage{soul}
\usepackage[dvipsnames]{xcolor}
\usepackage{amsmath}

\newcommand\scalemath[2]{\scalebox{#1}{\mbox{\ensuremath{\displaystyle #2}}}}

\usepackage{color}

\begin{document}
%
\title{Comparative Analysis of Terahertz Propagation Under Dust Storm Conditions on Mars and Earth}

\author{Lasantha~Thakshila~Wedage, Bernard~Butler\thanks{\it{Lasantha Thakshila Wedage and Bernard Butler are with the Walton Institute, South East Technological University, Ireland.}}, Sasitharan~Balasubramaniam\thanks{\it{Sasitharan Balasubramaniam and Mehmet C. Vuran are with the University of Nebraska-Lincoln, USA.}}, Yevgeni~Koucheryavy, \thanks{\it{Yevgeni Koucheryavy is with the Tampere University of Technology, Finland.}} \\Mehmet C. Vuran }

\maketitle

\begin{abstract}
Reliable Terahertz (THz) links are necessary for outdoor point-to-point communication with the exponential growth of wireless data traffic. This study presents a modified Monte Carlo simulation procedure for estimating THz link attenuation due to multiple scattering by dust particles on the THz beam propagation path. Scattering models are developed for beams through dust, based on Mie and Rayleigh approximations for corresponding frequencies for Earth (0.24 THz) and Mars (1.64 THz). The simulation results are compared, considering parameters such as the number of Monte-Carlo photon (MCP) packets, visibility, dust particle placement density along the beam, frequency, and distance between the transmitter and the receiver. Moreover, a channel capacity model was proposed, considering THz link attenuation due to dust storms, spreading loss and molecular absorption loss for Earth and Mars outdoor environments. Simulation results for Earth show that link attenuation increases with dust particle placement density, distance and frequency, and attenuation decreases with visibility. On Mars, similar results are obtained, except that the attenuation is variate around a constant value with the frequency increase. Channel capacity is estimated for Earth and Mars environments considering time and distance-dependent scenarios. Time windows that show a sudden drop of dust particles along the beam provide opportunities to communicate with high reliability. Moreover, increasing the distance between the transmitter and receiver severely reduces the channel capacity measurement in strong dust storm conditions in both environments. Our study has found that weak dust storms have relatively little effect on Mars, but much larger effects on Earth.
\end{abstract}


\IEEEpeerreviewmaketitle

\input{intro}

\section{Background}
\label{background}

\subsection{THz link behaviour in Dust Storms}
\label{behaviour}
 
THz signal attenuation due to the scattering loss caused by high dust particle density on the THz beam propagation path is the main concern of this study. Dust particle density on Mars is expected to be higher than on Earth because of the dusty atmosphere with low water vapour concentration. Mars dust consists of basalt and montmorillonite clay \cite{Nasa2002radio}. On the other hand, Earth dust consists of pollen, bacteria, smoke, ash, salt crystals from the ocean, and small amounts of dirt or various rocks, including sand. Moreover, during dust storm conditions on Mars, the effective radius of the dust particle varies from 1 to 4 microns with an effective variance of 0.2 – 0.4 \cite{Dustproperties}. However, on Earth, the effective radius varies between 1 and 150 microns \cite{Montecs,Nasa2002radio}. 

Many researchers investigated the THz \cite{Montecs, ep_value, dustAtt2020, su2012experimental} and lower frequency bands \cite{mmwave2009mathematical, li2010attenuation} attenuation due to the presence of dust particles on the beam propagation path. In \cite{Montecs}, Monte-Carlo simulation was used to calculate the transmittance of EM waves when they propagate through dust, considering multiple scattering effects for charged particles in 20 and 75 GHz frequencies. Hongxia et al. \cite{dustAtt2020} also studied the attenuation characteristics of THz waves subject to multiple scattering caused by dust storms in the Tengger desert, using the \emph{Mie} scattering approximation and Monte Carlo simulation. In addition, considering the Mie theory, Diao et al. \cite{ep_value} investigated THz wave attenuation due to heavy dust in the Martian atmosphere in the 0.1-1 THz frequency range and compared with Earth measurements. \cite{su2012experimental} investigated  attenuation at 0.625 THz caused by dust utilising an experimental setup and found that degradation of the THz link budget is minor due to dust, compared to that found using IR  beams with 1.5 $\mu m$ wavelength, and average attenuation of the THz link is proportional to the dust particle density. Moreover, Elshaikh  et al. \cite{mmwave2009mathematical} developed a mathematical model to characterise the microwave attenuation due to dust, considering parameters such as visibility, frequency, particle size and complex permittivity. Li et al. \cite{li2010attenuation} calculated the light scattering properties of partially charged dust particles utilising Mie scattering theory for various frequencies and found that for higher THz frequency EM waves, the attenuation effect of charge carried by sand particles can be ignored. Furthermore, \cite{2014rayleigh} presents the EM scattering properties of the small partially charged sand/dust particles, using the Rayleigh approximation, for microwave frequencies. 

\subsection{Atmospheric Condition Differences between Mars and Earth}
 When THz radio waves pass through the atmosphere, the signals experience attenuation due to many factors, which differ in their impact between Earth and Mars. This study focuses on point-to-point signal degradation in the lower part of the atmosphere (the troposphere) on Earth and Mars, when communicating antennas are placed 50 meters above the ground. Apart from improved line-of-sight properties, \cite{wedage2022path} shows that longer communication distances can be achieved on Mars because dust particle density decreases with height. The propagation medium in the troposphere of both planets includes gases, water vapour, clouds, fog, ice, dust, and assorted aerosols (haze), but the proportions vary. The impairment mechanisms include absorption, scattering, refraction, diffraction, multi-path, scintillation and Doppler shift. Impairment phenomena include fading, attenuation, depolarization, frequency broadening, and ray bending. However, this study considers only Line-of-Sight (LoS) transmission under dust storm scenarios through the troposphere of both Earth and Mars. It considers signal attenuation based on three factors: (1) free space path loss (which is the same for Earth and Mars), (2) molecular absorption due to atmospheric gases (which are different for Earth and Mars), and (3) scattering loss due to dust particles along the propagation path (Mars and Earth typically have different dust distributions). 

Free space path loss occurs due to misalignment between the transmitter and the receiver antennas. It is the same for both environments because it only depends on carrier frequency and distance. Molecular absorption loss plays a significant role on both planets. It measures the fraction of power loss (of the carrier wave) converted to kinetic energy due to molecular vibration when EM waves propagate through molecules of the atmosphere. Therefore, unlike spreading loss, molecular absorption loss depends on local atmospheric gas composition and density (see Table \ref{tab:AtmosphereComparison}), including carrier frequency and distance between the transmitter and the receiver. According to \cite{Rappaport2021}, certain frequencies of the THz spectrum, such as 183, 325, 380, 450, 550, and 760 GHz, suffer attenuation that is significantly greater than the free space propagation loss, due to water vapor absorption on Earth. However, the Martian atmosphere contains only about 1/1,000 as much water as Earth’s. Still, even this tiny amount can condense out, forming clouds that ride high in the atmosphere or swirl around the slopes of towering volcanoes \cite{Nasa2002radio}. This serious issue needs to be considered for vertical communication of Mars surface devices (Rovers, Habitats, etc.) and satellites. Since our study focuses on horizontal point-to-point communication, we do not need to consider upper atmospheric layer’s impact on THz signal transmission. Therefore, at these frequencies, we expect lower molecular absorption loss and higher channel capacity on Mars compared to Earth.

To the best of our knowledge, this is the first study that compares attenuation (at THz frequencies) due to dust storms on Earth and Mars, applying Monte Carlo simulation to the corresponding Mie and Rayleigh approximations. This paper also presents a channel capacity model that includes the effect of spreading, molecular absorption and dust scattering losses for sub-THz and THz links on Earth and Mars, respectively.

\section{THz beam propagation through a simulated 3D dust storm}
\label{3Dstorm}

This section discusses THz wave propagation through a randomly simulated 3D dust storm by simulating wind having both vertical (up-draught) and horizontal velocity components. This is used to calculate the number of dust particles on the beam path. The simulated dust storm (see Fig. \ref{fig:multiple scattering}) consists of a line source  starting at $X=0$ and a vertically upward movement of dust based on the vertex motion due to wind turbulence at $(6000,0,0)$. The line source dust storm in this study spreads for 8$m$ along the Y-axis $(-4 \leq Y \leq 4 )$. Such a line source is more realistic than a point source for dust storm simulation on both Earth and Mars.

MATLAB's wind package was used to simulate the storm and considered an exponential movement of dust along the positive X-axis coupled with strong wind in the same direction. Also, dust particles gradually precipitate from the atmosphere, when their weight exceeds upward forces. Dust particle movement of the point source dust storm downstream of the line source dust storm comprises both upward (point) wind and horizontal (line) wind, resulting in a vortex flow (a simulated whirlwind).

When counting the number of dust particles on the cone-shaped THz beam, we followed the following method. First, consider the THz beam starting at the point of $(0,0,h)$, where $h$ (50 $m$) is the transmitter antenna height and beam propagation direction is aligned with the positive X-axis. In this cone-shaped beam, the maximum radius of the impact area is calculated to be approximately 15 $cm$ for the corresponding transmitting distance of 10 $Km$. Since it is difficult to calculate the dust particle concentration on such a pencil-thin beam, we sub-divided the cone-shaped beam into $1 \times 10^6$ disks with 1 $cm$ distances between the disk centres (see Fig. \ref{fig:multiple scattering} (a)). Then we identified the position of each dust particle at 1 $m$ below the antenna height and recorded its position considering the nearest two disks. By looping over the position of each dust particle, and comparing the distances between the nearest disk centres and the particle position to centre distance, we encoded the position as being inside (1) or outside (0) of the THz beam for each particle. From this, we calculated the number of dust particles along the THz propagation path.    

Considering the dust particle size on Mars to vary between 0.5 to 4 microns \cite{Dustproperties,DPsize}, we simulate a scenario of sending a THz signal with the transmitter position (5500,0,50) and the receiver position (6500,0,50), which creates a 1000 $m$ distance between them while placing the point source vertex movement at (6000,0,0). As a result, we found that the average ratio of the number of generated dust particles along the line of propagation of the THz transmission beam is 0.0022872. Moreover, considering just the dust particles on the cone-shaped beam path, their density averaged 10.1832 (say, 10) particles per $cm^{3}$. Hence we can conclude that assuming the beam face area is 0.01 $cm^2$, the number of dust particles along the beam for a distance of 10 $m$ between the transmitter and the receiver is approximately 100 particles. However, in this random walk simulation process, it is difficult to control the effective radius of the dust particles. Therefore, depending on the dust particle size, scattering effects (which depend on the number and size of the particles) along the beam propagation path can vary.

\section{Modelling Monte Carlo Photon packets propagation through Dust particles}\label{MCPsim}
\begin{figure}
    \centering
    \includegraphics[width=\linewidth]{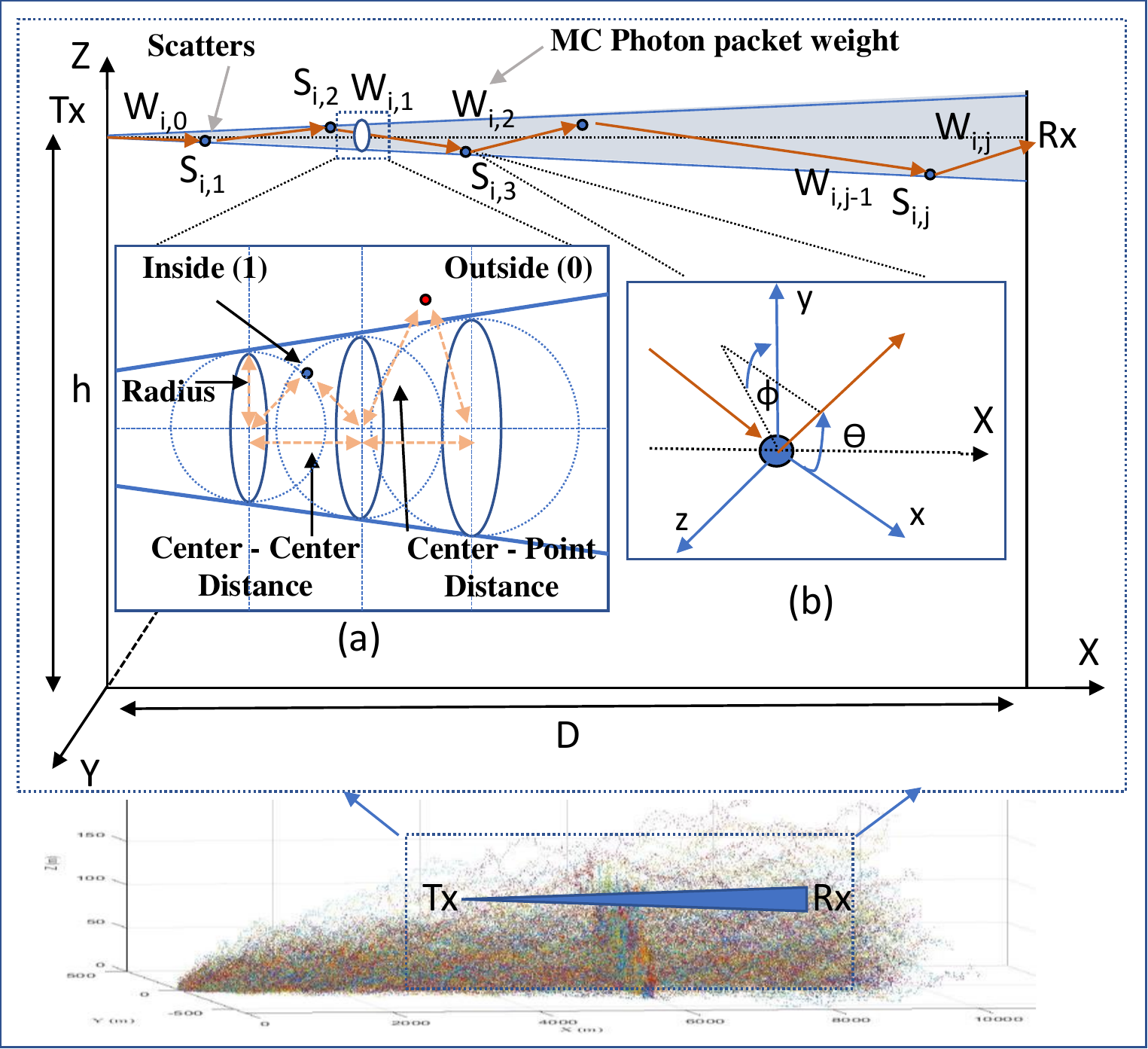}
    \caption{Multiple scattering processes of EMWs in a sand/dust storm with (a) the decision-making (in/out) method of dust particles from the beam and (b) the local coordinate system.}
    \label{fig:multiple scattering}
\end{figure}

In this section, we calculate the transmittance and the corresponding attenuation of the THz EM wave when it propagates through suspended dust particles. The initial intention was to consider the THz EM wave as a collection of photons. However, photon position and trajectory are not meaningful here \cite{mishchenko2009gustav} but collections of photons enable us to discretize the beam in a physically meaningful way. Monte Carlo simulation is used to estimate transmittance, where the incident plane EM wave is discretized as Monte Carlo photon (MCP) packets/units. Such photon packets provide an appropriate physical unit for discrete event simulation \cite{Montecs}. Each MCP packet is considered to be an equally divided portion of the energy weight of the EM wave field. In this simulation model, we assume that the particle number concentration is uniformly distributed throughout the THz beam area, and dust particles are randomly positioned. 

The intensity ($I$) of the incident THz EM wave can be expressed as $I_0=M.W$, where $M$ is the number of MCP packets per unit area per unit time and $W$ is the energy weight of each MCP. Here we suppose that MCP packet $i~(i=1,2,...,M)$ is randomly scattered by dust particles $j$ before it either exits the beam cone or reaches the receiver interface boundary at $X=D$. We assume that MCP packets enter from the point $(0,0,h)$ (which is height $h$ corresponding to the transmitter antenna height) (see Fig.\ref{fig:multiple scattering}) and are forward-scattered by scattering particles $S_{i,l}$ $(l=1,2,...,j)$ whose positions are denoted by $(x_{i,l},y_{i,l},z_{i,l})$, which are assumed random. Moreover, the algorithm will randomly select the number of scattering particles ($l$) that collide with an MCP packet from $j$ dust particles on the EM wave propagation path because each MCP packet will randomly change its propagation direction after every collision and dust particles are uniformly distributed along the beam path.

We suppose the initial energy weight of each MCP packet is $W=W_{i,0}=1$ and after scattering due to scattering particle $S_{i,l}$ is $W_{i,l}$. We also define a local coordinate system $xyz$ with the origin located at the scattering particle. So, we can define the propagation direction of the MCP packet $i$ with respect to the local coordinate system $xyz$, which is to the $x$ direction considering forward scattering. The propagation direction for the scattering direction of the MCP packet $i$ due to the impact with scattering particle $S_{i,l}$ can be expressed using the direction cosines $(\mu_{i,l}^x,\mu_{i,l}^y,\mu_{i,l}^z)$ in the global coordinate system $XYZ$, which is defined with the help of scattering (or deflection) angle $\theta$ and azimuth angle $\phi$ (see Fig. \ref{fig:multiple scattering} (b)). In each simulated collision, the scattering particle position $(x_{i,l},y_{i,l},z_{i,l})$ and the propagation angles ($\theta,\phi$) to calculate the direction cosines were calculated randomly for the global coordinate system, which will be explained in detail later.

Direction cosines can be calculated according to Fig.~\ref{fig:multiple scattering} (b) as follows,
\begin{subequations}
\label{eq1}
\begin{align}
\scalemath{0.8}{\mu_{i,l}^x} & \scalemath{0.8}{= -sin(\theta_{i,l})cos(\phi_{i,l}) \sqrt{1-(\mu_{i,l-1}^x)^2} + \mu_{i,l-1}^xcos(\theta_{i,l}) }\\ 
\scalemath{0.8}{\mu_{i,l}^y} & \scalemath{0.8}{= \dfrac{sin(\theta_{i,l})(\mu_{i,l-1}^y \mu_{i,l-1}^x cos(\phi_{i,l}) - \mu_{i,l-1}^z sin(\phi_{i,l}))}{ \sqrt{1-(\mu_{i,l-1}^x)^2}} + \mu_{i,l-1}^y cos(\theta_{i,l})}\\
\scalemath{0.8}{\mu_{i,l}^z} & \scalemath{0.8}{= \dfrac{sin(\theta_{i,l})(\mu_{i,l-1}^z \mu_{i,l-1}^x cos(\phi_{i,l}) + \mu_{i,l-1}^y sin(\phi_{i,l}))}{ \sqrt{1-(\mu_{i,l-1}^x)^2}} + \mu_{i,l-1}^z cos(\theta_{i,l})}.
\end{align}
\end{subequations}

If $|\mu_{i,l-1}^x|>0.99999$, then
\begin{subequations} 
\label{eq2}
\begin{align}
\mu_{i,l}^x & =\dfrac{\mu_{i,l-1}^x}{|\mu_{i,l-1}^x|}cos(\theta_{i,l})\\
\mu_{i,l}^y & = sin(\theta_{i,l})cos(\phi_{i,l})\\
\mu_{i,l}^z & = sin(\theta_{i,l})sin(\phi_{i,l})
\end{align}
\end{subequations}

In the initial stage, we suppose that the MCP packet $i$ enters the dust particle zone form $X=0$ at point (0,0,h) and propagates along the direction of $(\mu_{i,0}^x,\mu_{i,0}^y,\mu_{i,0}^z)=(1,0,0)$. The simulation process depends on uniformly-distributed, randomly generated numbers $\epsilon_{i,l},\nu_{i,l},$ and $\chi_{i,l} \sim Uniform(0,1)$, which are used to calculate the random variables $\Delta S_{i,l}, \theta_{i,l}$ and $\phi_{i,l}$.

 The $\Delta S_{i,l}$ is the travelling distance between the scattering particles $S_{i,l}$ and $S_{i,l-1}$ and defined as,
\begin{equation} \label{eq3}
\begin{split}
\Delta S_{i,l} & = \dfrac{-ln(\epsilon_{i,l})}{C_{ext}}\\
\end{split}
\end{equation}   
where $C_{ext}$ is the total extinction cross-section efficiency of spherical dust particles with radius $r$. The total extinction cross-section efficiency is expressed \cite{Montecs} as,  
 \begin{equation}
     \label{EarthExt}
     C_{ext}=\int_{r_{min}}^{r_{max}}{N_0 P(r)c_{ext}d(r)},
 \end{equation}
where $P(r)$ is the log-normal size distribution of dust particles for both Earth \cite{Earth_logN} and Mars \cite{ep_value} environments. Here, $N_0$ is the number of dust particles per unit volume, and it can be expressed as a function of visibility ($V_b$) \cite{ep_value}, which is represented as, 
\begin{equation}
       N_0 = \dfrac{15}{0.034744 V_b  \int_{0}^{2r_{max}}{\pi r^2 P(r)dr}}.
\end{equation}
On Earth, dust particle radius varies between 1 to 150 $\mu m$. Therefore, when considering the approximate equality of the effective diameter of the dust particles and the wavelength of the THz frequency utilised in this study, we can use the Mie approximation \cite{Mieapproximation} to infer the total extinction cross-section ($c_{ext}$) \cite{Nasa2002radio}, which is the sum of the absorption cross-section and scattering cross-section. The $c_{ext}$ is expressed by the Mie solution for spherical particles with dielectric constant $\epsilon$ \cite{mmwave2009mathematical} as, 
\begin{equation}
    \label{Mie}
    c_{ext} = \dfrac{k^3r\lambda^2}{2}(c_1 + c_2 (kr)^2 + c_3 (kr)^3)
\end{equation}
where,
\begin{subequations}
\begin{align}
c_1 & = \dfrac{6\epsilon''}{(\epsilon'+2)^2+ (\epsilon'')^2} \\
\scalemath{0.8}{c_2} &\scalemath{0.7}{ = \epsilon'' \Big\{ \dfrac{6}{5} \dfrac{7(\epsilon')^2 + 7(\epsilon'')^2 + 4\epsilon' -20}{[ (\epsilon'+2)^2+ (\epsilon'')^2 ]^2} \Big\} + \dfrac{1}{15} + \dfrac{5}{3[(2\epsilon'+3)^2+ 4(\epsilon'')^2]^2}}\\
\scalemath{0.8}{c_3} &\scalemath{0.7}{ = \dfrac{4}{3}\Big\{ \dfrac{(\epsilon'-1)^2 (\epsilon'+2) + [2(\epsilon'-1)(\epsilon'+2)-9]+(\epsilon'')^4}{[(\epsilon'+2)^2+ (\epsilon'')^2]^2} \Big\}}.
\end{align}
\end{subequations}
The complex refractive index of dry dust particles on Earth can be expressed as $\sqrt{3 + \imath.\frac{18.256}{f}}$ \cite{dustAtt2020}, where $f$ is the frequency and $\imath$ is the imaginary unit $\sqrt{-1}$. 

On the other hand, the total extinction cross-section of the dust particles on Mars can be evaluated utilising the Rayleigh approximations \cite{2014rayleigh} because the effective diameter of the dust particles on Mars (1-8 $\mu m$) is less than one-tenth of the wavelength of the frequency \cite{rayleigh2016} used in this study. Therefore, the total extinction cross-section can be expressed as,
\begin{equation}
\scriptstyle{
    \label{MarsExt}
   \scalemath{0.8}{ c_{ext} = \dfrac{8}{3}\pi k^4 r^6 \Big|\dfrac{\epsilon_r - 1}{\epsilon_r + 2} \Big|^2 + 12 \pi k \epsilon^{''}_r r^3 \dfrac{1}{|\epsilon_r + 2|^2} + 
    \dfrac{\pi}{6} \dfrac{k^4 r^6 \sigma^2}{E^{2}_0\epsilon^{2}_0} |\epsilon_r - 1|^2}
    }
\end{equation}
where $\epsilon_r$ is the relative permittivity  \cite{2014rayleigh}.
Moreover, we can take the complex refractive index of dust particles on Mars as $1.52 + 0.01i$ \cite{ep_value, wedage2022path} corresponding to the radius range (0.5-4 $\mu m$) used in this study.

The scattering angle, $\theta_{i,l}$, due to the $i^{th}$ MCP packet impact with scatter $l$ is represented as,
\begin{equation} 
\label{eq4}
\begin{split}
\theta_{i,l} & =  \left\{
  \begin{array}{l} 
      \scalemath{0.8}{cos^{-1} \Big\{ \dfrac{1}{2g}  \Big((1+g^2) - \Big(\dfrac{1-g^2}{1-g+2g\nu_{i,l}}\Big)^2\Big)\Big\} \hspace{1mm} \text{for} \hspace{2mm} g \neq 0} \\
      \scalemath{0.8}{cos^{-1} (2 \nu_{i,l} - 1) \hspace{1mm} \text{for} \hspace{2mm} g = 0}
     \end{array}
     \right.
\end{split}
\end{equation}
where $\phi_{i,l}$ is the azimuth angle due to the same impact and 
\begin{equation} \label{eq5}
\begin{split}
\phi_{i,l} & =2 \pi \chi_{i,l}\\
\end{split}
\end{equation}
and $g=<cos(\theta)> \in [0,1]$ is the asymmetry factor (here, $g=0$ refers to the isotropic scattering and 1 refers to forward direct scattering). We assume $g$ varies uniformly between 0.5 and 1 for our simulations which is the average value for the direct forward scattering. 

After calculating the random variables $\epsilon_{i,l},\nu_{i,l},$ and $\chi_{i,l}$, we can determine the (random) position of the scattering particle $S_{i,j}$ ($X_{i,l}, Y_{i,l}, Z_{i,l}$) in the global coordinate system utilising the equations (\ref{eq1}), (\ref{eq2}) and (\ref{eq3}) as below. The position of the scattering particles can be expressed as,
\begin{equation} \label{eq6}
\begin{split}
X_{i,l} & = X_{i,l-1} + \Delta S_{i,l} \hspace{0.8mm} \mu_{i,l-1}^x\\
Y_{i,l} & = Y_{i,l-1} + \Delta S_{i,l}  \hspace{0.8mm} \mu_{i,l-1}^y\\
Z_{i,l} & = Z_{i,l-1} + \Delta S_{i,l} \hspace{0.8mm}  \mu_{i,l-1}^z\\
\end{split}
\end{equation}
Successful transmittance occurs only for MCP packets that reached the receiver interface at a distance of $D$ from the transmitter. If $X_{i,l} >= D$, this means that $S_{i,l-1}$ is the last scattering particle encountered by the MCP packet $i$ before it leaves the region boundary ($X=D$) from the receiving interface. Therefore, we can stop the simulation process and go to the next MCP packet ($i+1$) after calculating its current energy weight ($W_{i,l}$). The energy weight follows the Beer-Lambert law, which determines how $W_{i,l-1}$ is related to the $W_{i,l}$ \cite{weight,Montecs}. Thus, the energy weight of an MCP packet after collision with a scattering particle can be expressed as, 
\begin{equation} \label{eq7}
\begin{split}
W_{i,l} & = W_{i,l-1} \hspace{0.8mm} exp{\Big\{ \dfrac{-C_{ext} (X_{i,l}-X_{i,l-1})}{\mu_{i,l-1}^x} \Big\}}.\\
\end{split}
\end{equation}
If $X_{i,l} < D$, this means MCP packet $i$ is unable to reach the receiver interface after impacting with $l$ scatters. From this point, we focus our interest more on the energy weight of the MCP packet $i$ and calculate the energy weight of the MCP packet $W_{i,l}$ using eq. (\ref{eq7}). In this instance, we assume that the initial energy weight of the MCP packet is a unit (i.e., $W_{i,0}=1$), where we define a threshold ($\epsilon_t$) value of $1 \times 10^{-5}$ to consider as the minimum energy weight that an MCP packet can take after $l$ impacts with the scatters. If the energy weight of an MCP packet does not exceed this minimum threshold (i.e., $W_{i,l} < \epsilon_t $), the packet is assumed not to reach the receiver and is recorded as such. Therefore, we can set $j=l$ and $W_{i,l}=0$.

Based on the calculation of energy weights of the MCP packets that reached the receiver interface, we can calculate the transmittance of the THz EM wave by,
\begin{equation} \label{eq8}
\begin{split}
T_{MS} &  = \dfrac{\sum_{i=1}^{M}{W_{i,l} \hspace{0.8mm} exp \Big \{ \dfrac{ -C_{ext} (D-X_{i,l})}{\mu_{i,l}^x}\Big\}}}{I_0}. \\
\end{split}
\end{equation}   
From our calculations, we noticed that eq. (\ref{eq8}) does not converge to a finite value for every simulation. Therefore, we assume the transmittance to be zero when it is divergent and set the simulation to the next Monte-Carlo process. Based on the transmittance measurements at the end of this procedure, we can calculate the specific attenuation $A_{MS}$ in  $(dB/m)$ according to \cite{Montecs}, as,
\begin{equation} \label{eq9}
\begin{split}
 A_{MS} & = \dfrac{-4.343 \hspace{0.8mm} ln(T_{MS})}{D}.\\
\end{split}
\end{equation} 
\section{THz Channel Capacity}
\label{channelcap}
To evaluate the channel capacity in the THz band, we can decompose the received signal as a sum of the sub-bands, where each sub-band channel is narrow and has a flat-band response \cite{multi}.

The $i^{th}$ frequency sub-band is defined as $\Delta f_i = f_{i+1} - f_i$ with power $P_i$ under the constraint $\sum_{i=1}^{N_B}{P_i \leq P_t}$, where $N_B$ refers to the total number of sub-bands, and $P_t$ stands for the total transmit power. In the $i^{th}$ narrow-band, the sub-band capacity, $C_i$, is expressed in \cite{multi} as,
\begin{equation}
\label{cm1}
    C_i = \Delta{f_i} \log{\Big( 1+ \dfrac{|h_{LoS}|^2 P_i}{\Delta{f_i S_D(f_i)}}\Big)},
\end{equation}
where $S_D$ is the power spectral density of the additive white Gaussian noise (AWGN) and $h_{LoS}$ is the frequency-dependent channel response for attenuation due to dust particles including spreading loss and molecular absorption loss due to gas molecules on the LoS signal propagation path. According to \cite{multi}, $h_{LoS}$ can be expressed as,
    \begin{equation}
        \label{h_los}
        h_{LoS} (\tau) = \alpha_{Los} \delta(\tau -\tau_{LoS}).
    \end{equation}
where $\alpha_{Los}$ refers to the attenuation and $\tau_{LoS}$ refers to the propagation delay due to dust particles and gas molecules on the signal propagation path and $\tau_{LoS} = \dfrac{d_{LoS}}{c}$. Where $d_{LoS}$ is the signal travelling distance through dust, which is $D$ in this study, because we calculate the transmittance using the MCP packets that reached the fixed receiver interface. Also, the power spectral density of AWGN can be expressed as $S_D(\tau) = \frac{n_0}{2} \delta(\tau)$ and, in the frequency domain, $PSD(S_D(f))$= $\frac{n_0}{2}$. Utilizing the Wiener-Khinchin theorem, the frequency-dependent channel response for LoS attenuation can be expressed as \cite{multi},
\begin{equation}
\label{cm3}
    h_{LoS} (\tau)= |H_{LoS} (f)| \delta(\tau -\tau_{LoS}).
\end{equation}
The free space direct ray or LoS channel transfer function, $H_{LoS}$, consists of the spreading loss function ($H_{Spr}$), the molecular absorption loss function ($H_{Abs}$), and scattering loss function due to dust particles ($H_{Dust}$), which can be expressed as,
\begin{equation}
\label{H_los}
    H_{LoS} (f)= H_{Spr}.H_{Abs}.H_{Dust} e^{-j 2 \pi f \tau_{LoS}}.
\end{equation}
The free space path loss or the spreading loss ($PL_{Spr}$) measures the fraction of power lost by a beam with frequency $f$ over a distance $D$ in a vacuum, and it can be expressed according to \cite{jornet2011channel} as,
\begin{equation}
\label{pl_spr}
        PL_{spr} = \Big(\frac{4 \pi D f}{c} \Big)^2,
    \end{equation}
where $c$ is the speed of light in the medium. Thus, according to \cite{multi} the corresponding channel transfer function for the spreading loss can be expressed as,
\begin{equation}
\label{H_spr}
        H_{spr} = (PL_{spr})^{-1/2}=\Big(\frac{c}{4 \pi D f} \Big).
\end{equation}
The molecular absorption loss measures the fraction of power converted to kinetic energy due to molecular vibration when EM waves propagate through molecules in the atmosphere. Thus, when transmitting frequency $f$ through a homogeneous medium between a transmitter and receiver at a distance $D$, the molecular absorption loss is obtained with the help of the Beer-Lambert law \cite{jornet2011channel}, which is represented as,
    \begin{equation}
    \label{pl_abs}
        PL_{abs} = e^{k(f)D},
    \end{equation}
where $k(f)=\sum_{i,g}^{}{k^{i}_{g}(f)}$ and $k^{i}_{g}(f)$ is the monochromatic absorption coefficient of the $i^{th}$ isotopologue of $g^{th}$ gas at frequency $f$. When calculating the absorption coefficient, we consider water vapour and nine other gases for Earth and six gases for Mars (see Table \ref{tab:AtmosphereComparison}), except for Argon. This allows us to consider the vastly different gas concentrations between the two planets. The monochromatic absorption coefficient for each isotopologue of a particular gas in the Martian and Earth atmosphere at frequency $f$ is provided in \cite{Hitran2016},
    \begin{equation}
        k^{i}_{g}(f) = S^{i}_{g}(T)F^{i}(f),
    \end{equation}
where $S^{i}_{g}(T)$ is the line intensity at temperature $T$ (210K for Mars) referenced to the temperature 296K of the $i^{th}$ isotopologue of $g^{th}$ gas, which can be easily calculated using the high-resolution transmission (HITRAN) molecular spectroscopic data. Where, $F^{i}$ is the spectral line shape function at frequency $f$. 
In the lower atmosphere on Earth, pressure broadening of spectral lines dominates the line shape and a Lorentz profile can be assumed as the line shape function and it is given by \cite{Hitran2016},
\begin{equation}
\label{F_L}
        F^{i}_L(f)=\dfrac{1}{\pi}\dfrac{\gamma(p,T)}{\gamma(p,T)^2 + [ f - (f^{i}_{g} + \delta(P_{ref}) P)]^2}
    \end{equation}
where $f^{i}_{g}$ is the resonant frequency for the isotopologue $i$ of gas $g$, $\gamma(P,T)$ is the Lorentzian (pressure-broadened) HWHM for a gas at pressure $P$ (atm), temperature $T$ (K), and $\delta(P_{ref})$ is the pressure shift at reference pressure ($P_{ref}$= 1 $atm$).

Since Doppler-broadening dominates the line shape in low-pressure environments such as Martian environment, a Gaussian profile can be assumed as the line shape function, and it is given by,
    \begin{equation}
    \label{F_G}
        F^{i}_G(f)=\sqrt{\dfrac{\ln{2}}{\pi {\alpha^{i}_D}^2}}\exp{\Bigg(-\dfrac{(f-f^{i}_{g})^2 \ln{2}}{{\alpha^{i}_D}^2}\Bigg)}\;,
    \end{equation}
where $\alpha^{i}_D$  is the Doppler broadening half-width,
    \begin{equation}
        \alpha^{i}_D = \dfrac{f^{i}_{g}}{c}\sqrt{\dfrac{2 N_A k_B T \ln{2}}{M^{i}}},
    \end{equation}
    where $M^{i}$ is the molar mass of isotopologues which can be obtained from the HITRAN database \cite{Hitran2016}, and $N_A$ and $k_B$ are the Avogadro and Boltzmann constants.
    
Thus, according to \cite{multi} the corresponding channel transfer function for the molecular absorption loss due to the gas molecules on the atmosphere can be express as,   
\begin{equation}
    \label{H_abs}
        H_{abs} = (PL_{Abs})^{-1/2}= e^{-\frac{1}{2}{k(f)D}},
    \end{equation}
Finally, $H_{Dust}$ is the transfer function for the attenuation due to dust particles that can be expressed as following the relationship between the transfer functions and the attenuation functions for spreading loss and molecular absorption loss, respectively, as well as utilising the dust attenuation function in eq. (\ref{eq9}) in dB, which is represented as,
\begin{equation}
\label{cm4}
    H_{dust}= \dfrac{1}{\sqrt{10^{-0.4343 \ln{(T_{MS})}}}}. 
\end{equation}
Therefore, substituting the functions derived for $H_{Spr}$, $H_{Abs}$, and $H_{Dust}$, to eq. (\ref{H_los}), we can calculate the channel transfer function. Moreover, in this study, we consider one narrow frequency band for each environment, which is 0.22 - 0.24 THz for Earth and 1.64 - 1.67 THz for Mars. Therefore, we can suppose that $P_i = P_t$ and the eq. (\ref{cm1}) can be rewritten as,
\begin{equation}
\label{cm2}
    C = \Delta{f} \log{\Big( 1+ \dfrac{|h_{LoS}|^2 P_t}{\Delta{f S_D(f)}}\Big)}.
\end{equation}

\begin{figure}[t]
    \centering
    \subfigure[Earth: Transmittance.]{\includegraphics[width=0.24\textwidth]{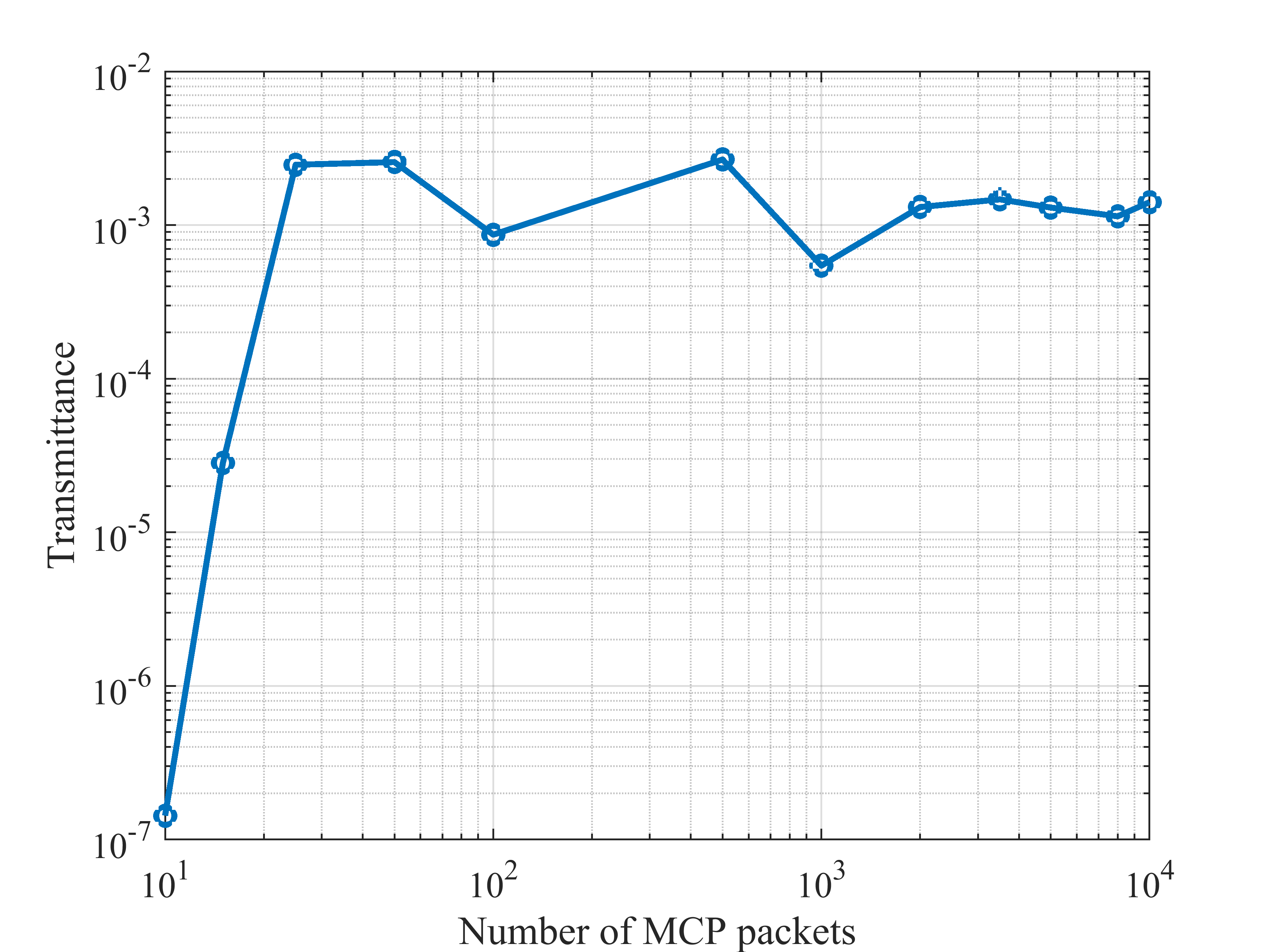}} 
    \subfigure[Earth: Attenuation ($dB/m$).]{\includegraphics[width=0.24\textwidth]{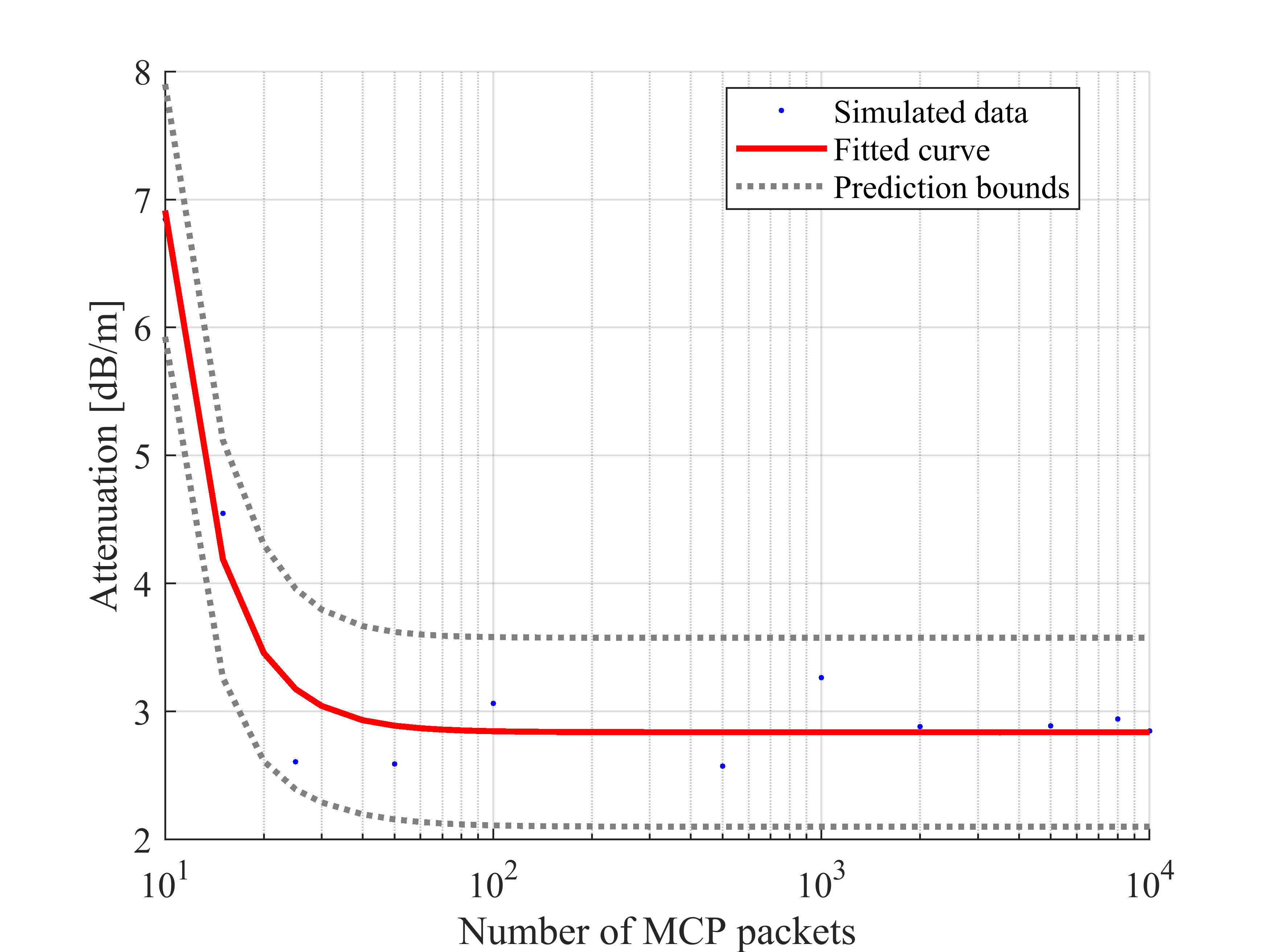}}
    \subfigure[Martian: Transmittance.]{\includegraphics[width=0.24\textwidth]{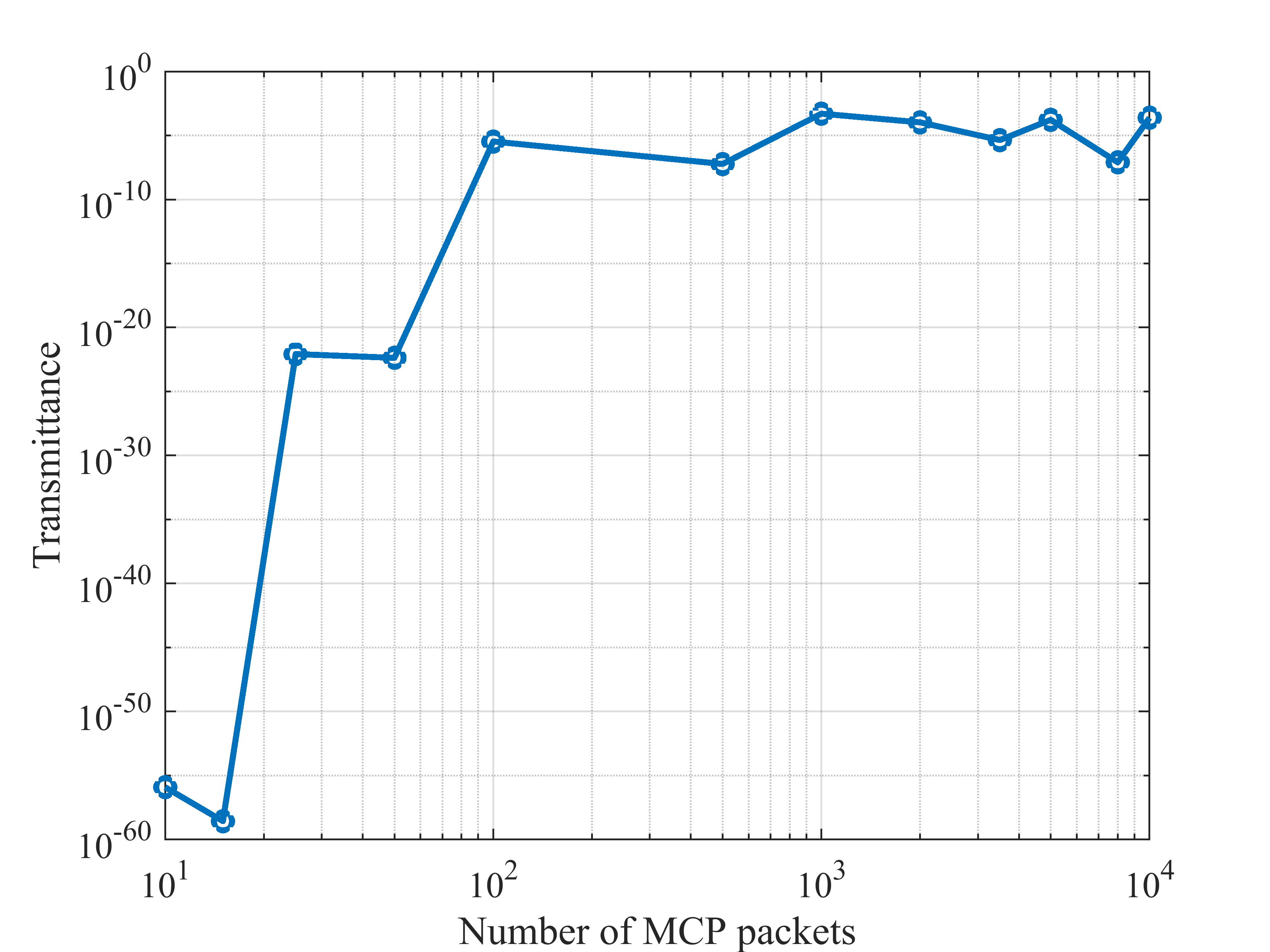}} 
    \subfigure[Martian: Attenuation ($dB/m$).]{\includegraphics[width=0.24\textwidth]{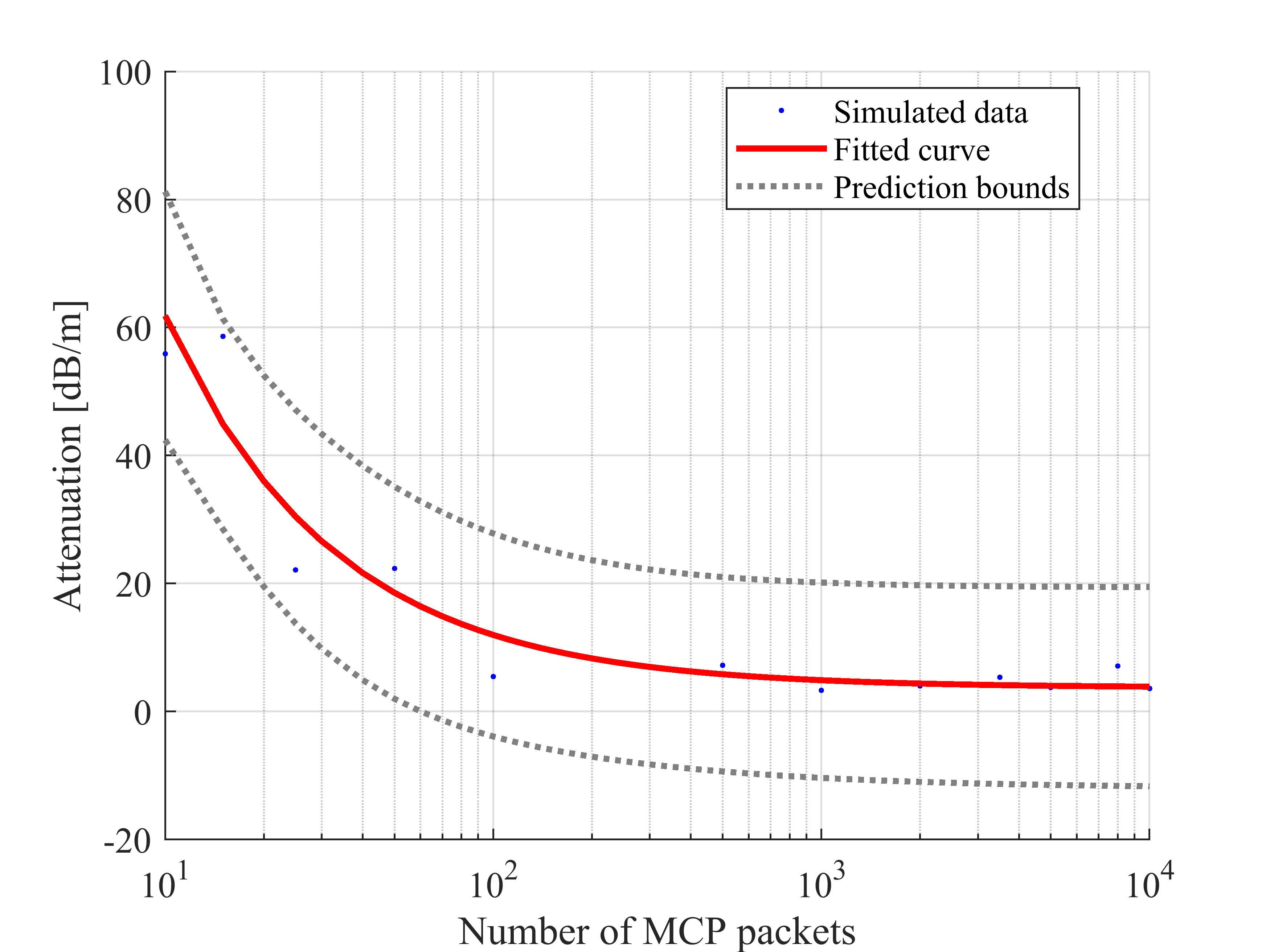}}
    \caption{Simulation measurements of a) the transmittance and b) the attenuation for a THz beam of 0.24 THz and 1.64 THz frequency for Earth and Mars by varying MCP packets from 10 to 10000 while fixing dust particle number (100/10000) on the propagation path, visibility and the distance (10 $m$) between transmitter and the receiver.}
    \label{fig:MCP}
\end{figure}

\section{Results and Discussion}
\label{RnD}

\subsection{Transmittance and Attenuation measurements through Monte-Carlo Simulations}

This subsection presents the simulation results for the transmittance and attenuation measurements of the THz link and the generated estimation models for the THz attenuation due to dust particles on the beam propagation path by varying parameters such as the MCP packets, visibility, dust particle number, the distance between transmitter and the receiver, and EM frequency. When simulating data for a targeted parameter, we have kept the other parameters constant to make the interpretation easy (see Table \ref{tab:Channel conditions on Earth and Mars}). Thus, we have chosen a frequency of 1.64 THz as the constant frequency for Mars \cite{wedage2022path}, and 0.24 THz frequency for Earth \cite{0.24THz_koenig2013wireless,survey2019} corresponding to low molecular absorption and high transmission distance. However, it is crucial to consider molecular absorption on Mars, even though it has a thin atmosphere with very low water vapour concentration. Moreover, we have considered 100 dust particles on the beam propagation path for a 10 $m$ fixed distance between the transmitter and receiver, as explained in section \ref{3Dstorm} for simulations on Earth. It is unrealistic to consider the same amount of dust particles for the Mars simulations because of the tiny particle sizes on Mars. Therefore, considering the blockage that this dust particle creates and the proportional relationship between blockage area and dust particle radius, we consider 10000 dust particles for Mars corresponding to 100 dust particles on Earth. When selecting the number of MCP packets, we considered 10000 packets in this study. 
\begin{table}[!ht]
\caption{Channel conditions and simulation settings on Earth and Mars.}
    \centering
    \label{tab:Channel conditions on Earth and Mars}
    \begin{tabular}{|l|l|l|}
    \hline
        Parameter & Earth & Mars \\
        \hline
        Frequency & 0.24 THz & 1.64 THz \\
        MCP packets & $10^4$ & $10^4$ \\
        Dust Density & $10^2$ per 10$m$ & $10^4$ for 10 $m$ \\
        Dust radius & 1--150 microns & 0.5--4 microns \\
        Dust size distribution & log-Normal & log-Normal \\
        Antenna height & 50 $m$ & 50 $m$ \\
        Approximation & Mie & Rayleigh \\
        Distance & 1 - 200 $m$ & 1 - 200 $m$ \\
        Temperature & 288 K & 210 K \\
        Surface Pressure & 1013 $mb$ & 6.1 $mb$ \\
        Surface density & 1.29 $Kg/m^3$ & 0.02 $Kg/m^3$ \\
        \hline
    \end{tabular}
\end{table}

Figure \ref{fig:MCP} illustrate the simulated data for the transmittance and attenuation measurements using the simulation setup explained in section \ref{MCPsim} for Earth and Mars environments by varying the number of MCP packets from 10 to 10000, while keeping the other parameters constant. As we can see from the figures, the transmittance measurements for both Earth and Mars environments (see Fig. \ref{fig:MCP} (a) and (c)) are increasing rapidly when the MCP packets increase at the beginning up to 100. After that, it converges to a particular value corresponding to each environment. Furthermore, attenuation measurements decrease following a power function for both Earth and Mars environments (see Fig. \ref{fig:MCP} (b) and (d)) and converge approximately to a value of 3.6 $dB/m$ and 2.8 $dB/m$, respectively. In addition, the fitted power function for attenuation against the MCP packets ($N_{MCP}$) can be expressed as $\text{Attenuation ($dB/m$)} = 2145 N_{MCP}^{-2.721 } + 2.839$ for Earth and $\text{Attenuation ($dB/m$)} = 410 N_{MCP}^{-0.8485} + 3.727$ for Mars.

\begin{figure}[!htb]
    \centering
    \subfigure[Earth: Transmittance.]{\includegraphics[width=0.24\textwidth]{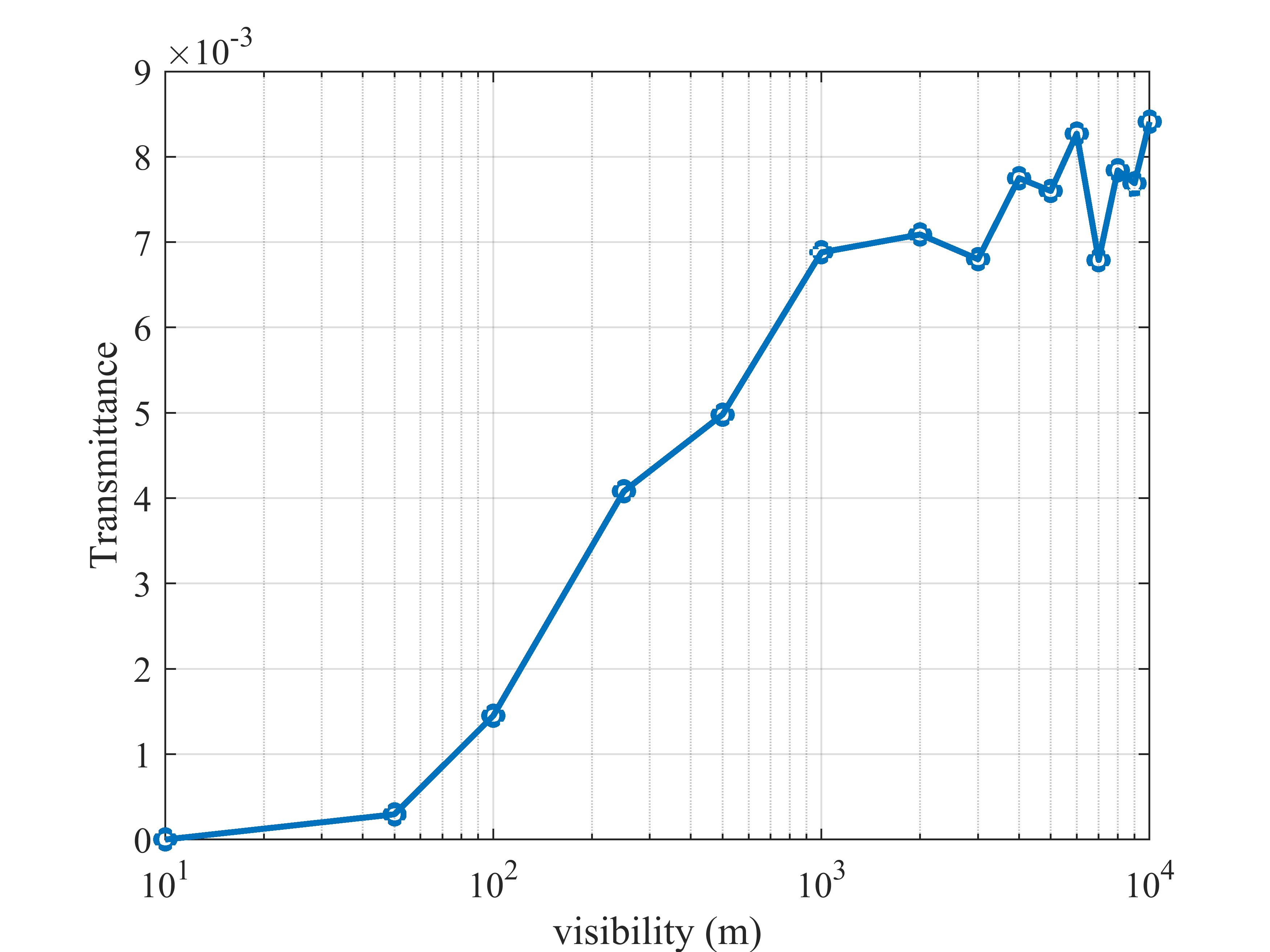}} 
    \subfigure[Earth: Attenuation ($dB/m$).]{\includegraphics[width=0.24\textwidth]{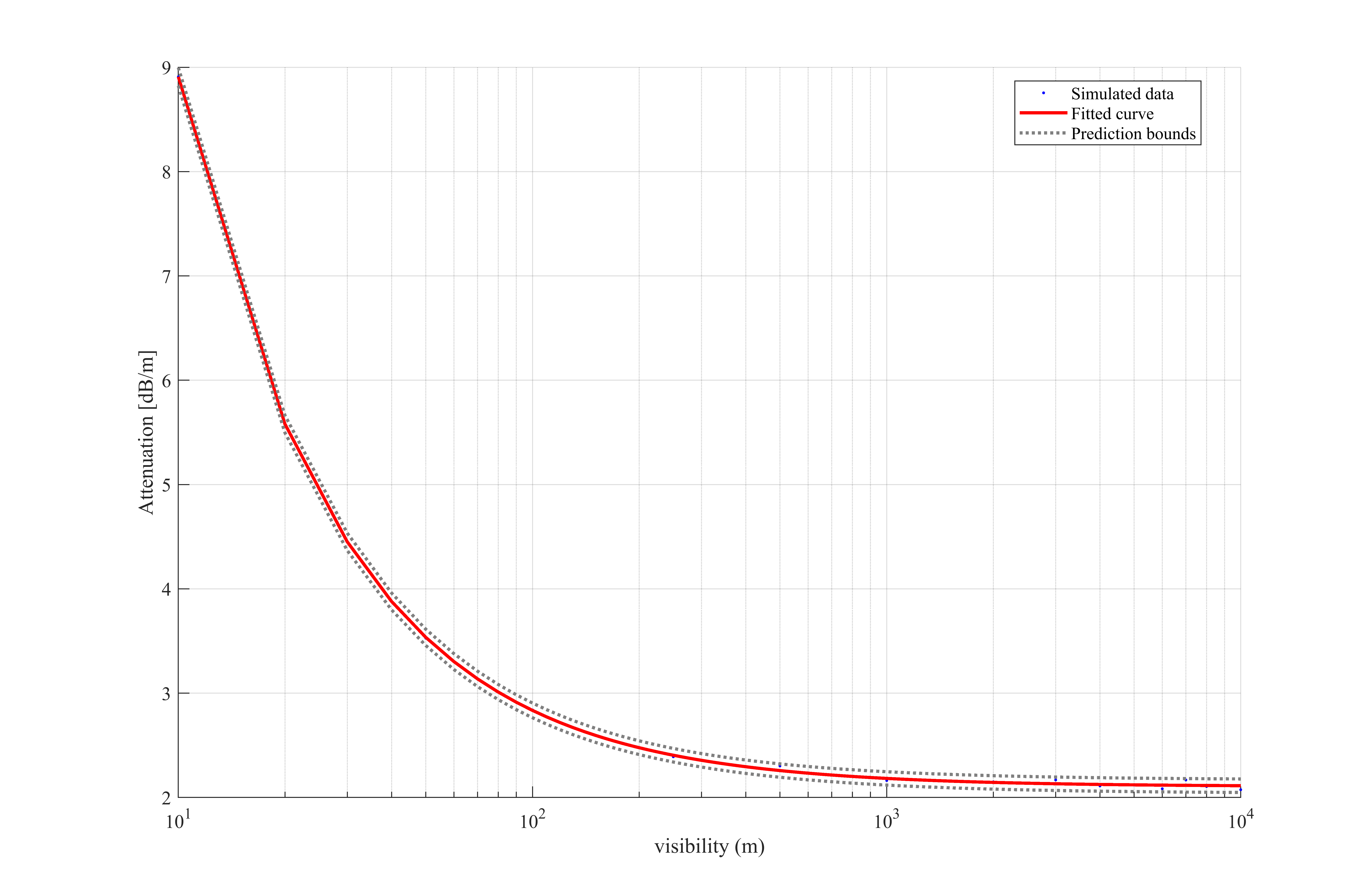}}
    \subfigure[Martian: Transmittance.]{\includegraphics[width=0.24\textwidth]{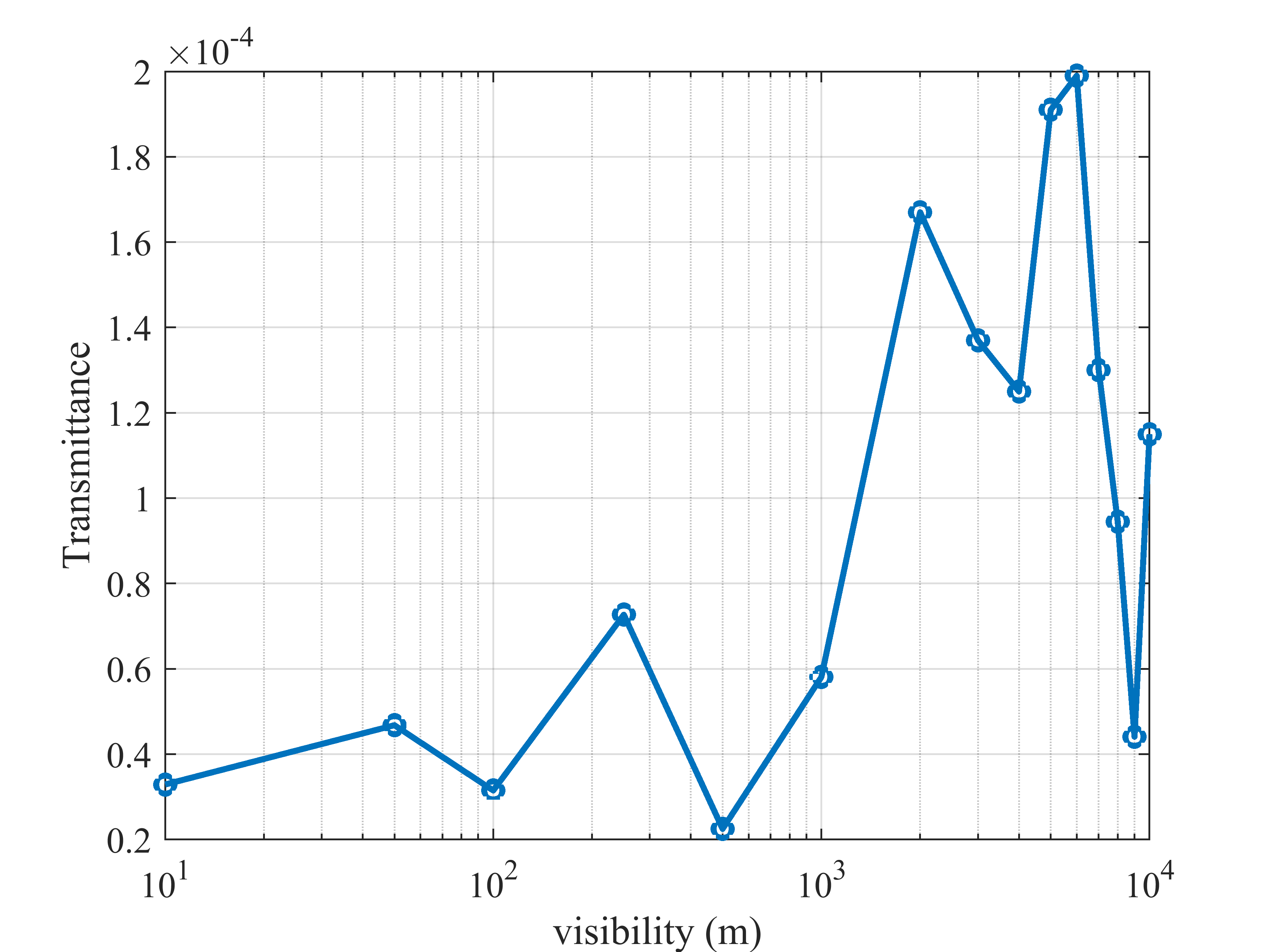}} 
    \subfigure[Martian: Attenuation ($dB/m$).]{\includegraphics[width=0.24\textwidth]{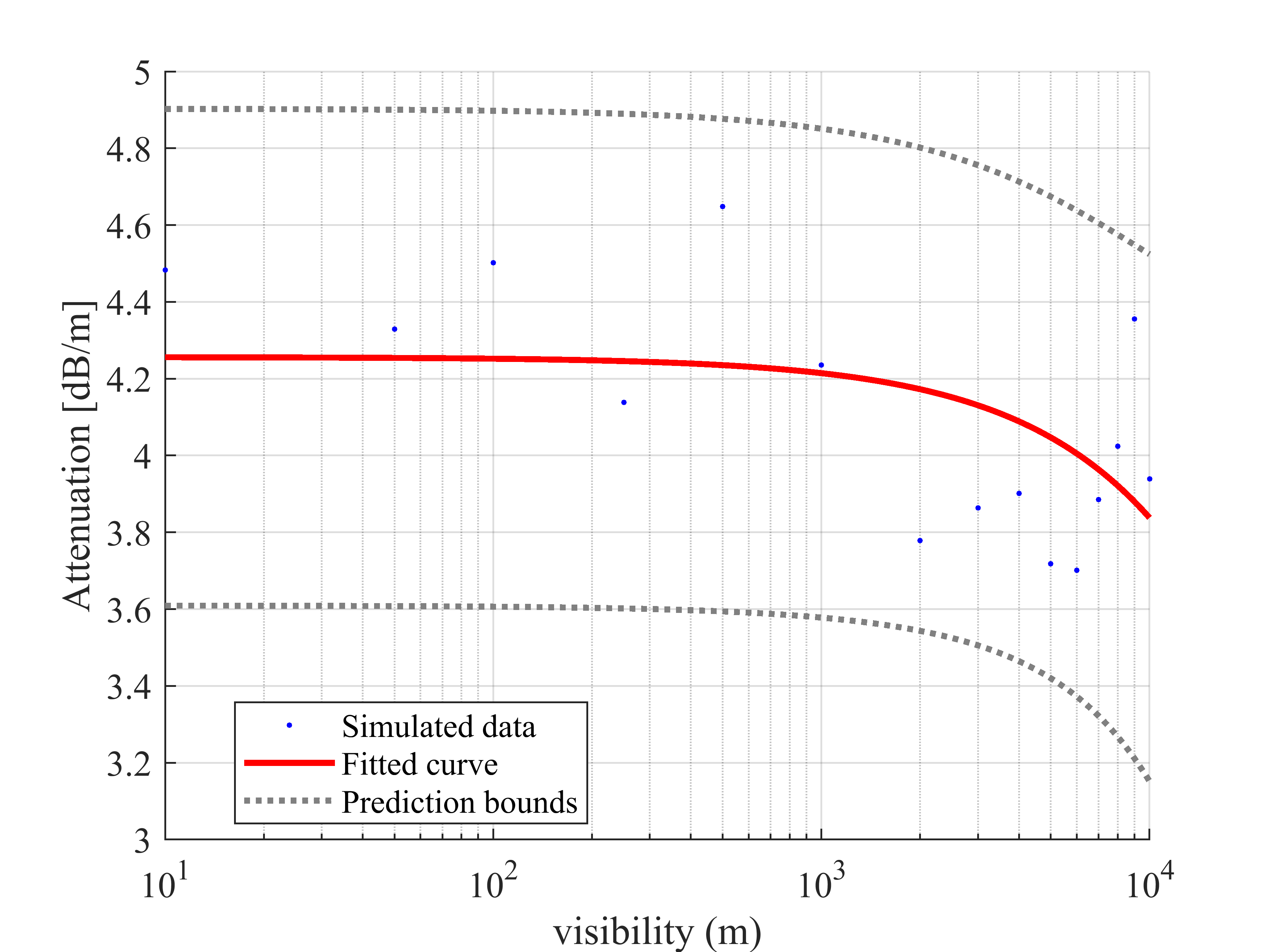}}
    \caption{Simulation measurements of a) the transmittance and b) the attenuation for a THz beam of 0.24 THz and 1.64 THz frequency for Earth and Mars, respectively, by varying the visibility from 10 to 10000 $m$ while fixing MCP packets (10000) and the distance (10 $m$) between transmitter and the receiver.}
    \label{fig:Visi}
\end{figure}

Next, we investigated the effect of visibility on the transmittance and attenuation measurement for Earth and Mars environments by varying the parameter values from 10 to 10000 (see Fig. \ref{fig:Visi}). Generally, when the visibility increases between the transmitter and the receiver, we will see fewer dust particles on the beam propagation path with a high distance variance between the particles. Therefore, we can expect high transmittance and low attenuation measurements when the visibility increases. According to Fig. \ref{fig:Visi} (a), the transmittance measurements of the Earth's environment are increasing dramatically, with the visibility and attenuation measurements (see Fig. \ref{fig:Visi} (b)) decreasing following a power function as expected. Moreover, the attenuation measurements approximately converge to  2.1 $dB/m$ value with an increase of visibility near 10000 $m$, which we can consider as clear sky condition. The fitted power function for the attenuation against the visibility ($V$) can be expressed as $\text{Attenuation ($dB/m$)} = 63.41 V^{-0.9694} + 2.105$ for the Earth environment. On the other hand, the transmittance measurements for the Mars environment (see Fig. \ref{fig:Visi} (c)) show an increasing trend with visibility. However, transmittance measurements for some visibility parameter values diverge from the trend because, according to our simulation process, each MCP packet will randomly select the scatters. Therefore, even if we have low dust density on the beam propagation path with high visibility, the transmittance can be low due to high collision with dust particles. Corresponding to the transmittance measurements, we can see a slight linear decrease in attenuation on Mars (see Fig. \ref{fig:Visi} (d)) with increased visibility. The fitted linear function can be expressed as $\text{Attenuation ($dB/m$)} = -4.184 \times 10^{-5} V + 4.256$.  

\begin{figure}[!htb]
    \centering
    \subfigure[Earth: Transmittance.]{\includegraphics[width=0.24\textwidth]{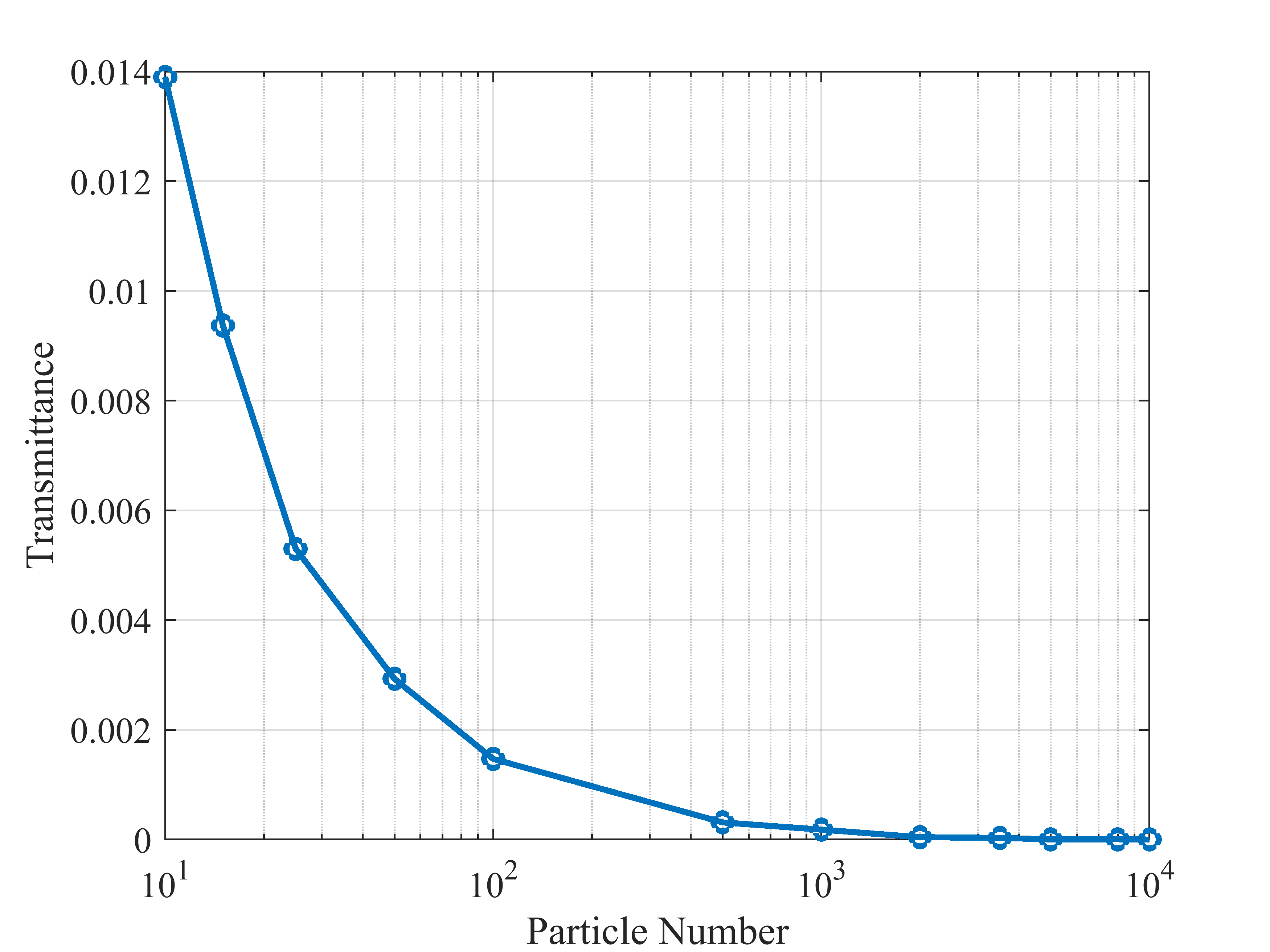}} 
    \subfigure[Earth: Attenuation ($dB/m$).]{\includegraphics[width=0.24\textwidth]{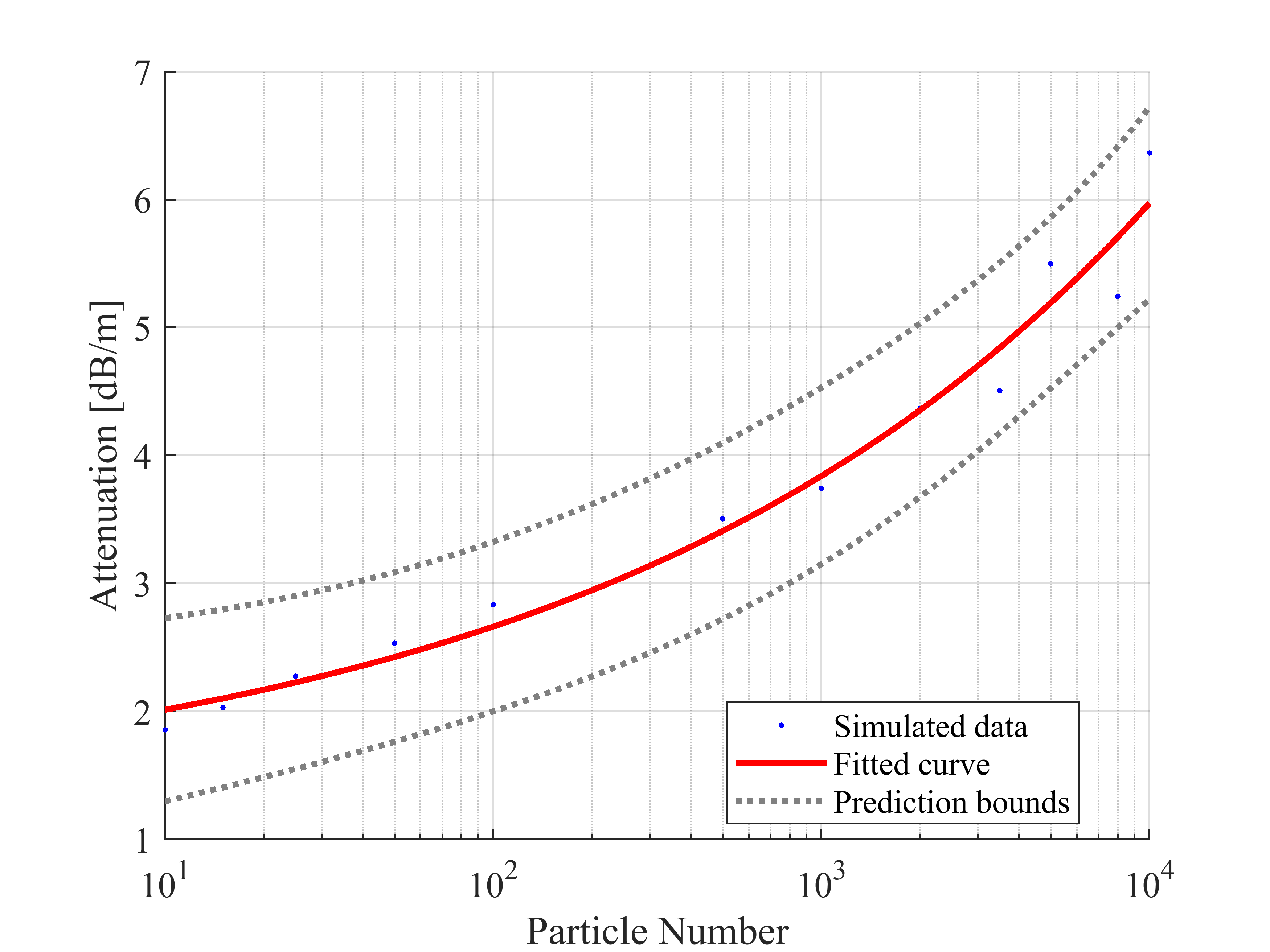}}
    \subfigure[Martian: Transmittance.]{\includegraphics[width=0.24\textwidth]{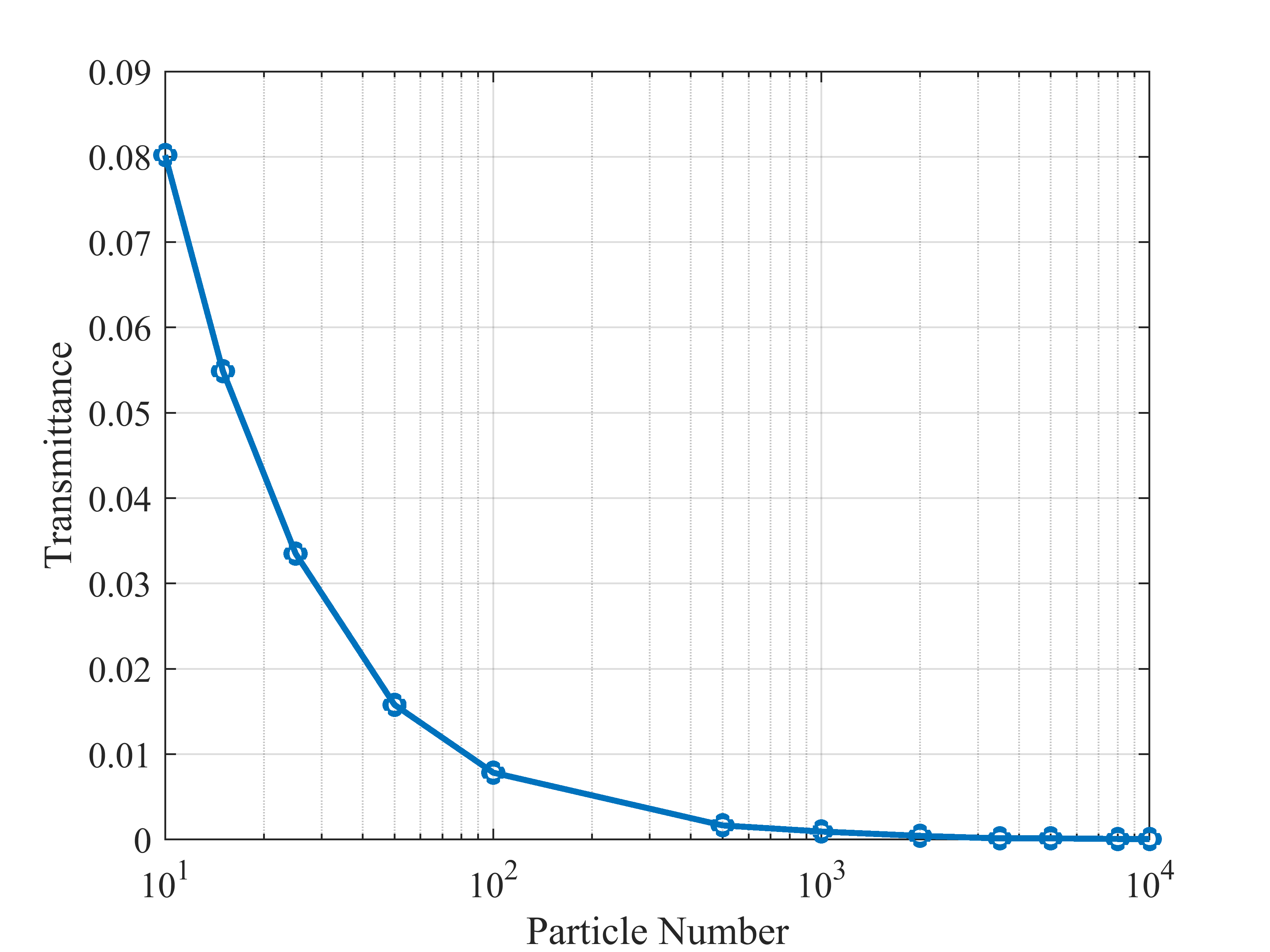}} 
    \subfigure[Martian: Attenuation ($dB/m$).]{\includegraphics[width=0.24\textwidth]{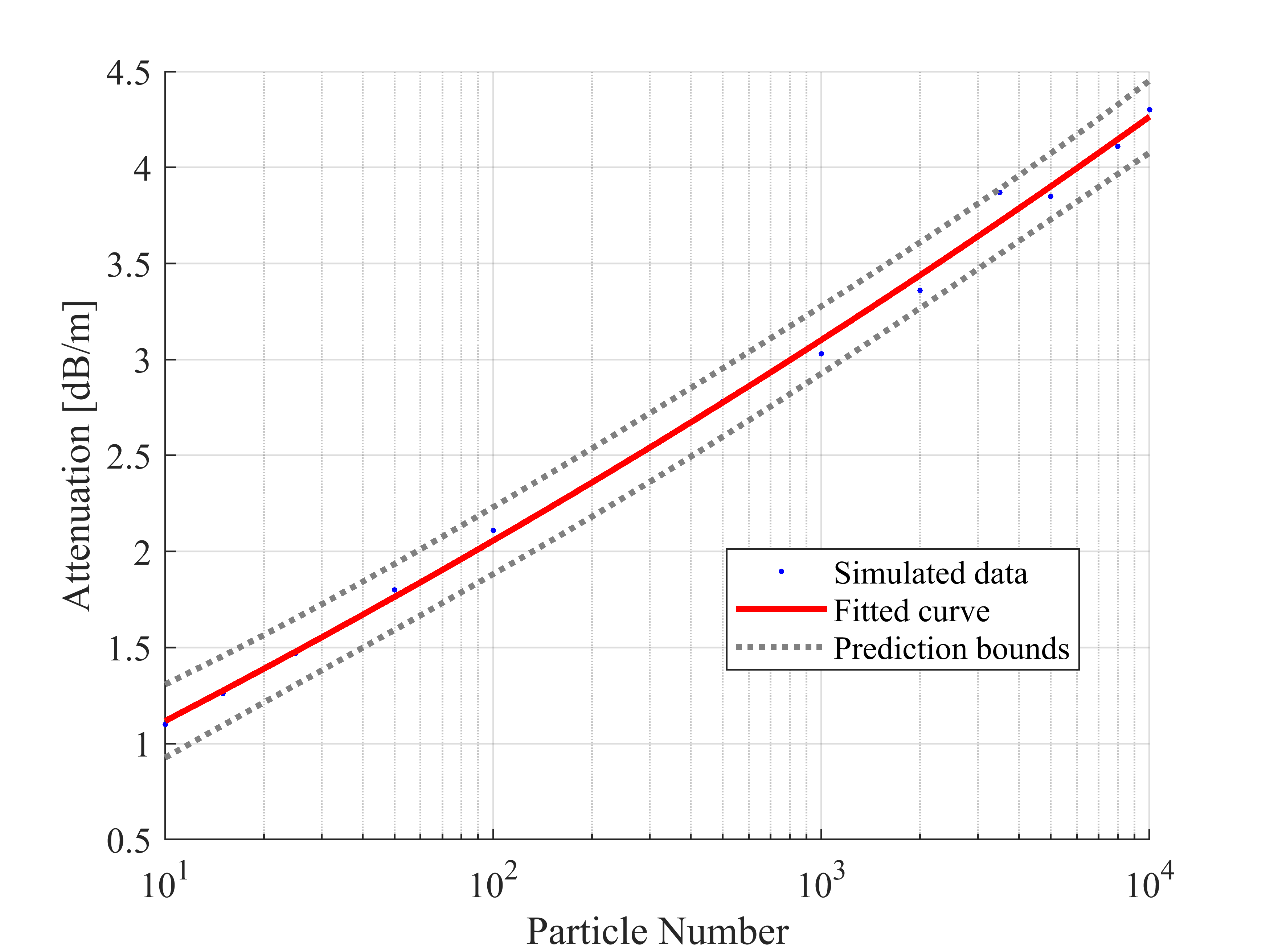}}
    \caption{Simulation measurements of a) the transmittance and b) the attenuation for a THz beam of 0.24 THz and 1.64 THz frequency for Earth and Mars, respectively, by varying the dust particle number on the beam propagation path from 10 to 10000 while fixing the number of MCP packets (10000) and the distance (10 $m$) between transmitter and the receiver.}
    \label{fig:PN}
\end{figure}

The dust particle density can vary unpredictably with the wind in a dust storm on Earth and Mars. There can be time windows with very low and high dust particles on the beam propagation path, which will be perfect for transmission. To investigate the effect of dust particle count, we have measured the transmittance and attenuation for Earth and Mars environments by varying dust particle numbers from 10 (very low) to 10000 (very high). As shown in Fig. \ref{fig:PN} (a) and (c), the transmittance measurement drops dramatically to near zero with the increase of dust particles for both environments. This rapid transmittance drop happens due to the high amount of scatters on the THz beam propagation path that each MCP packet should randomly collide.    Moreover, the attenuation measurements (see Fig. \ref{fig:PN} (b) and (d)) increased rapidly following a power function for both environments. The fitted power function for attenuation against the dust particle number ($D_{PN}$) can be expressed as $\text{Attenuation ($dB/m$)} = 0.4423 D_{PN}^{0.2579} + 1.213$ for Earth and $\text{Attenuation ($dB/m$)} = 7.534 D_{PN}^{0.04617} -7.262$ for Mars. Furthermore, as we can notice, attenuation measurements do not converge to a particular value as in previous cases when increasing the number of dust particles on the beam propagation path. Therefore, we can expect a communication blackout in a regional/global dust storm situation which will boost the dust particle density in the communication area.  

\begin{figure}[!htb]
    \centering
    \subfigure[Earth: Transmittance.]{\includegraphics[width=0.24\textwidth]{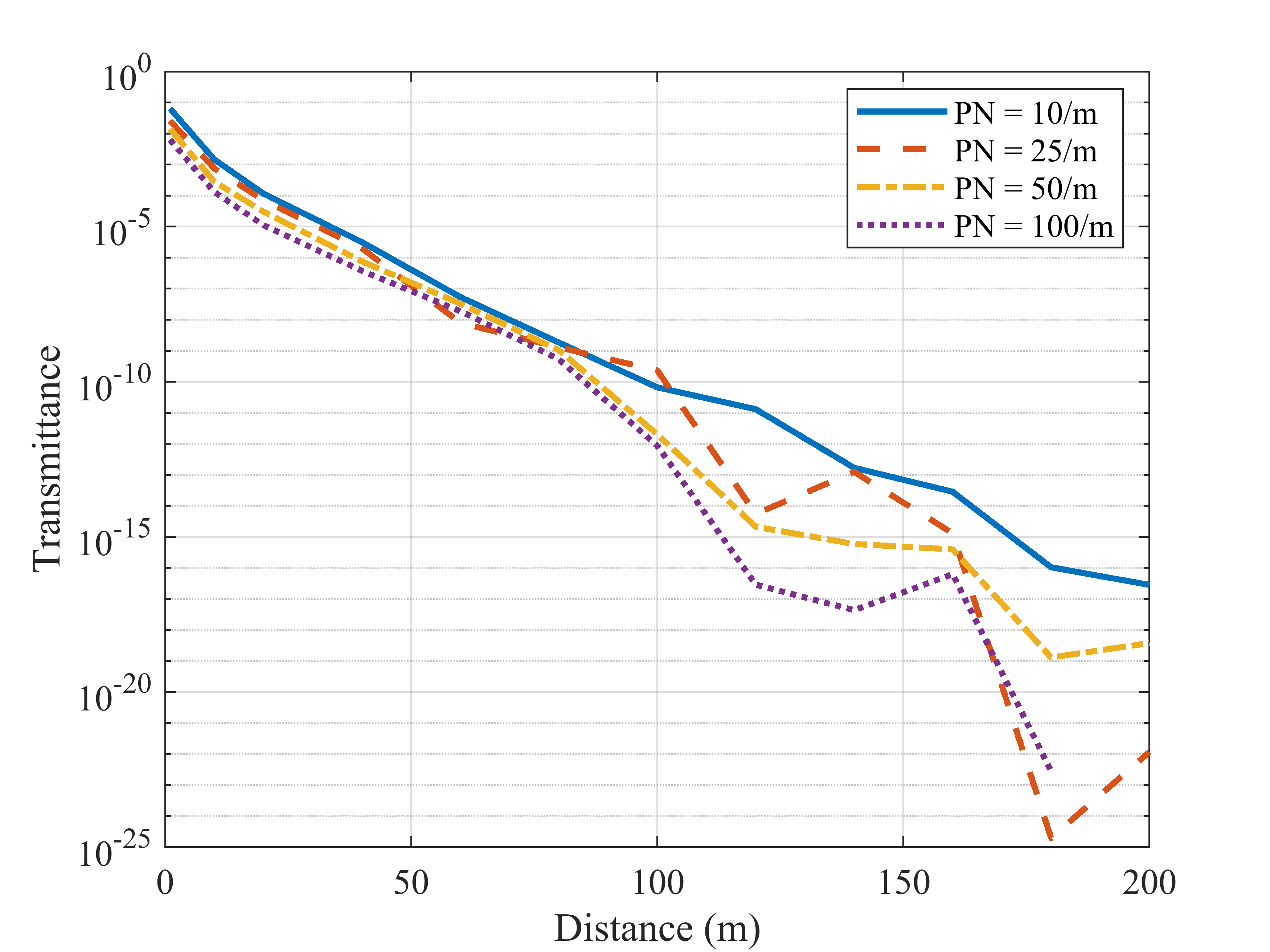}} 
    \subfigure[Earth: Attenuation (dB).]{\includegraphics[width=0.24\textwidth]{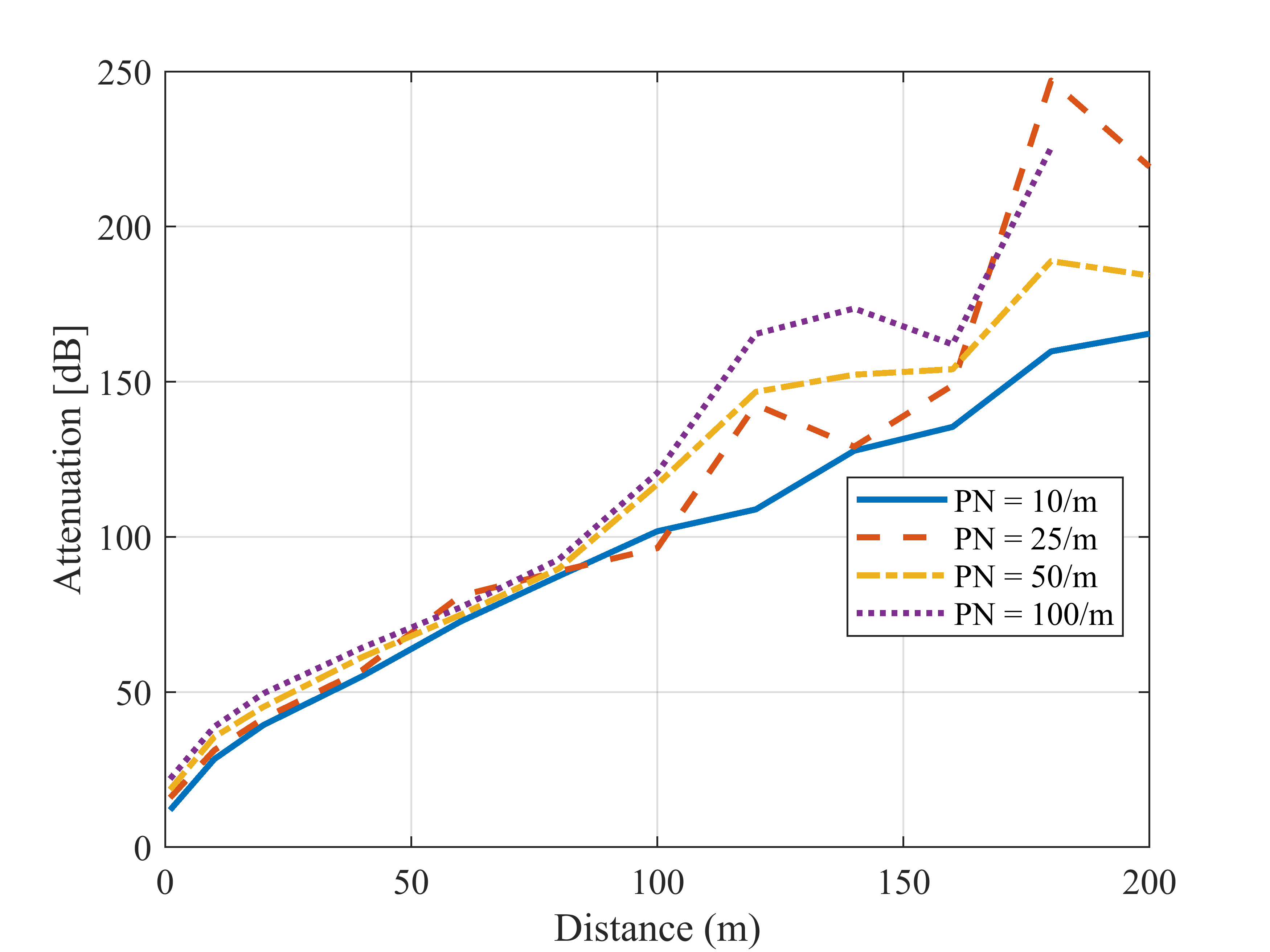}}
    \subfigure[Martian: Transmittance.]{\includegraphics[width=0.24\textwidth]{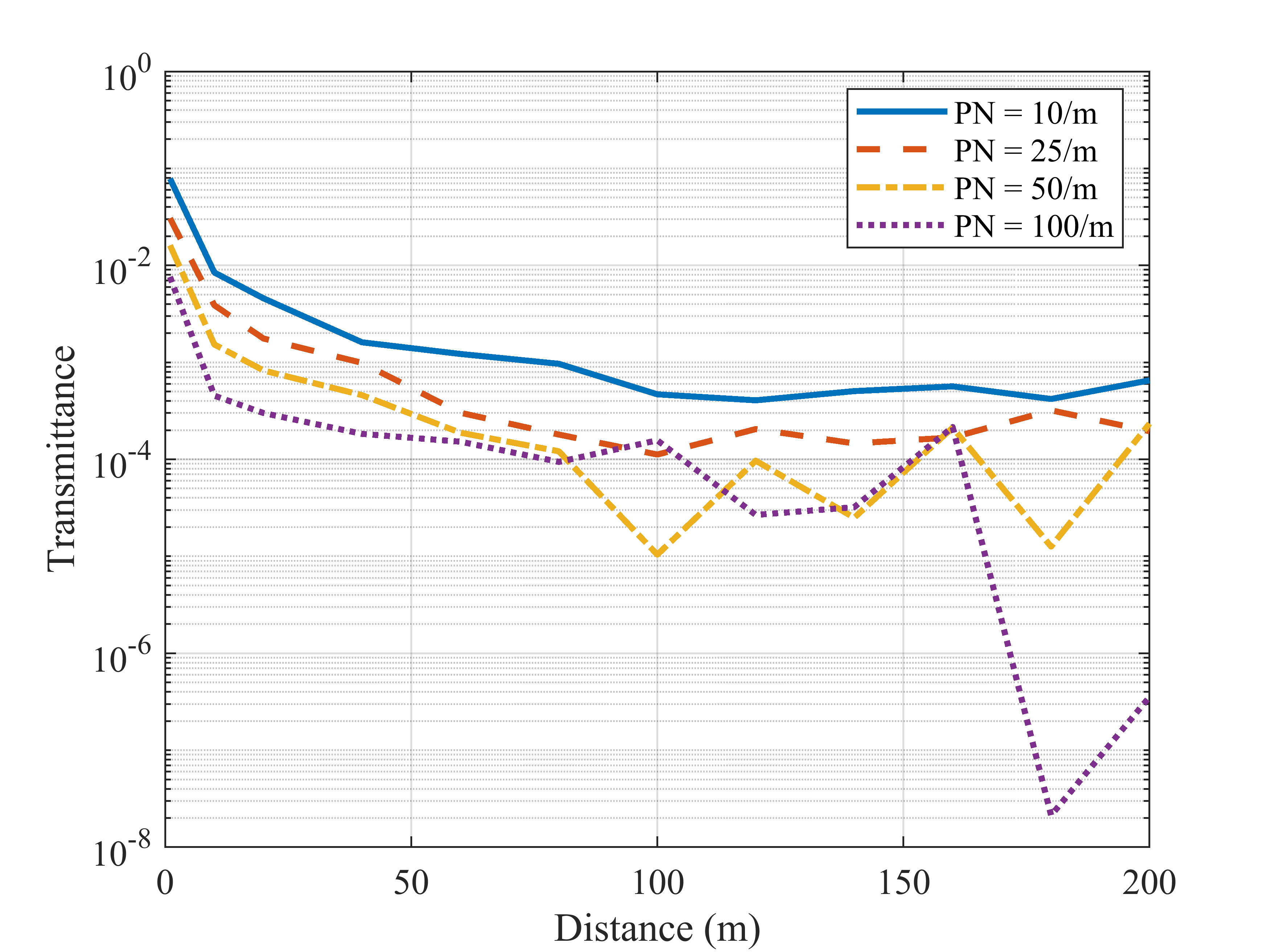}} 
    \subfigure[Martian: Attenuation (dB).]{\includegraphics[width=0.24\textwidth]{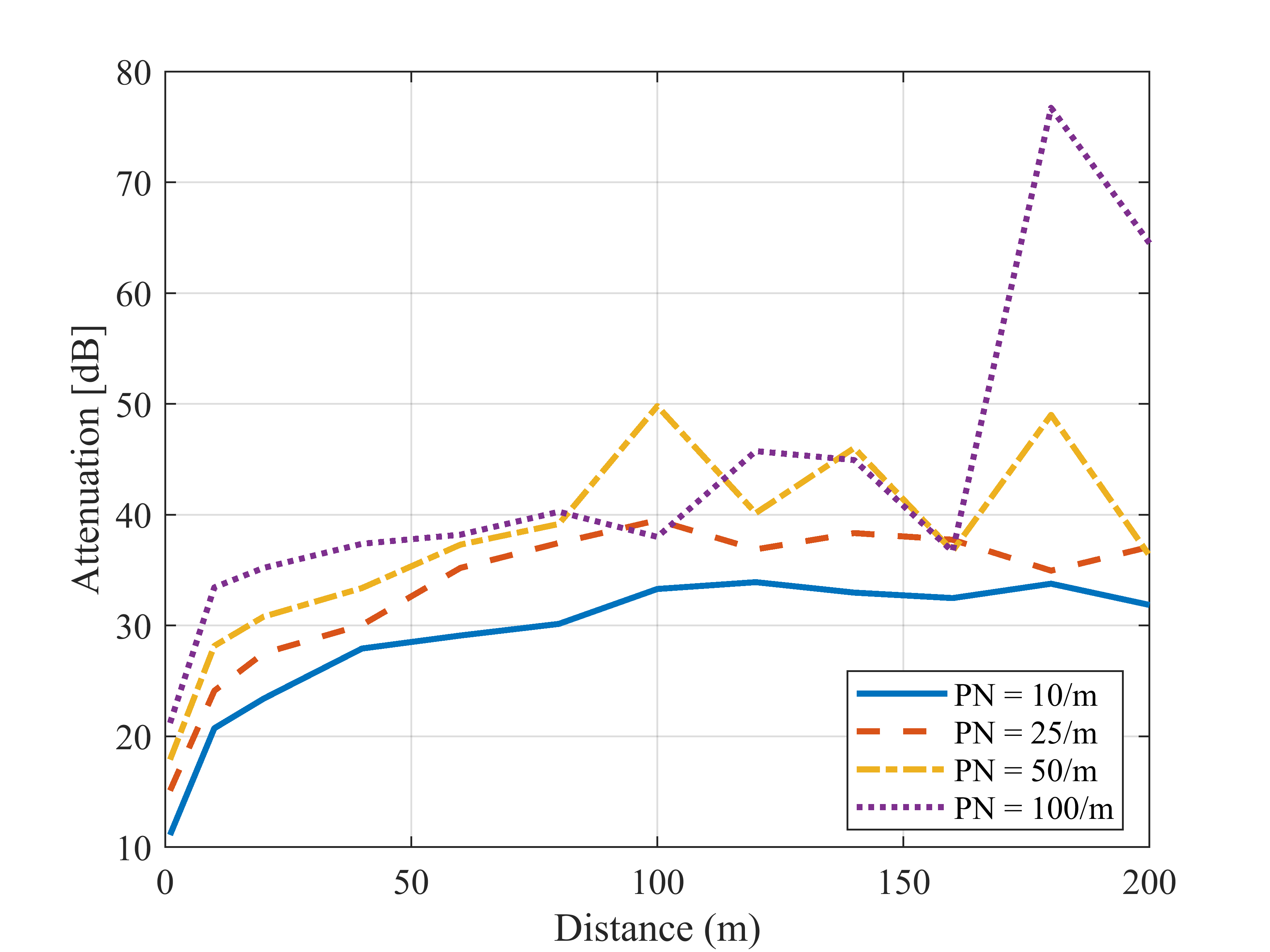}}
    \caption{Simulation measurements of a) the transmittance and b) the attenuation for a THz beam of 0.24 THz and 1.64 THz frequency for Earth and Mars, respectively, by varying the distance between transmitter and the receiver from 1 to 200 $m$ for different particle number densities of 10/$m$, 25/$m$, 50/$m$ and 100/$m$ on the beam propagation path while fixing MCP packets (10000).}
    \label{fig:PN_vary}
\end{figure} 

As mentioned above, dust density on the beam propagation path can vary significantly on Earth and Mars due to the unpredictable wind, temperature and pressure behaviour. Therefore, to conduct more realistic simulations, we have investigated the effect of distance between the transmitter and the receiver when we have various dust densities that vary with the distance. Dust density usually measures the number of dust particles per unit volume. However, in this study, we define it as the number of dust particles per meter for simplicity because the THz beam is assumed to be cone shape, and its face area is very tiny. This means that if we consider 10 dust particles per meter (10/$m$) for 100 $m$, we assume that 1000 dust particles are uniformly distributed on the beam propagation path. As we can see in Fig. \ref{fig:PN_vary} (a), the transmittance measurements on Earth drop dramatically with the distance, and when increasing the dust density, the transmittance measurements reaches near zero rapidly beyond 100 $m$. However, on Mars (see Fig. \ref{fig:PN_vary} (c)), the transmittance measurements decreases slightly compared to Earth. Moreover, we can clearly see that the transmittance measurements for 25/$m$ dust density are significantly lower than the dust density at 10/$m$, as expected. However, we can not see much difference between the transmittance measurements for 50/$m$ and 100/$m$ dust densities up to 150 $m$. On the other hand, attenuation measurements (see Fig. \ref{fig:PN_vary} (b) and (d)) advance with the increase of distance and dust density for both environments corresponding to transmittance measurements.

\begin{figure}[!htb]
    \centering
    \subfigure[Earth: Transmittance.]{\includegraphics[width=0.24\textwidth]{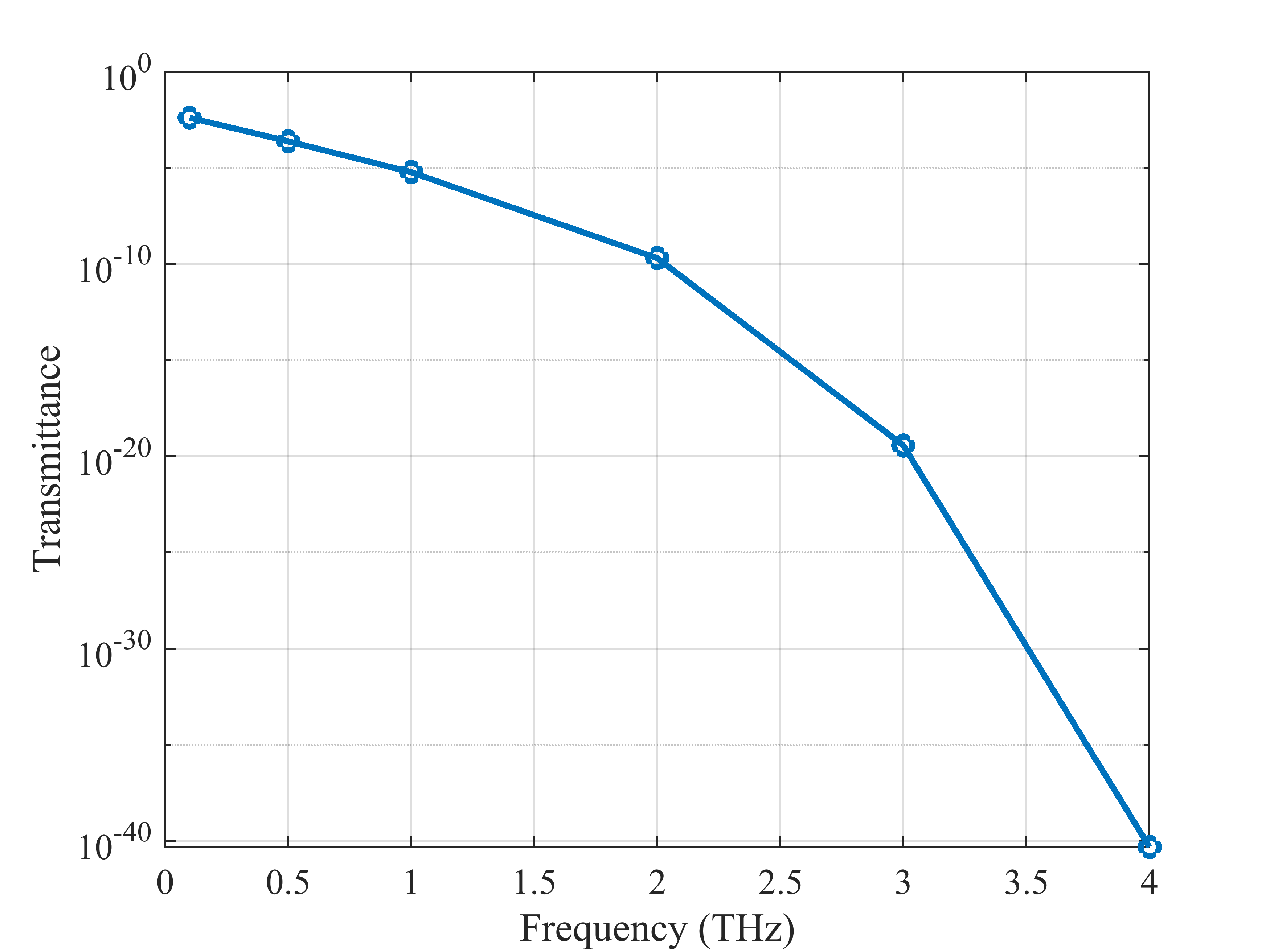}} 
    \subfigure[Earth: Attenuation ($dB/m$).]{\includegraphics[width=0.24\textwidth]{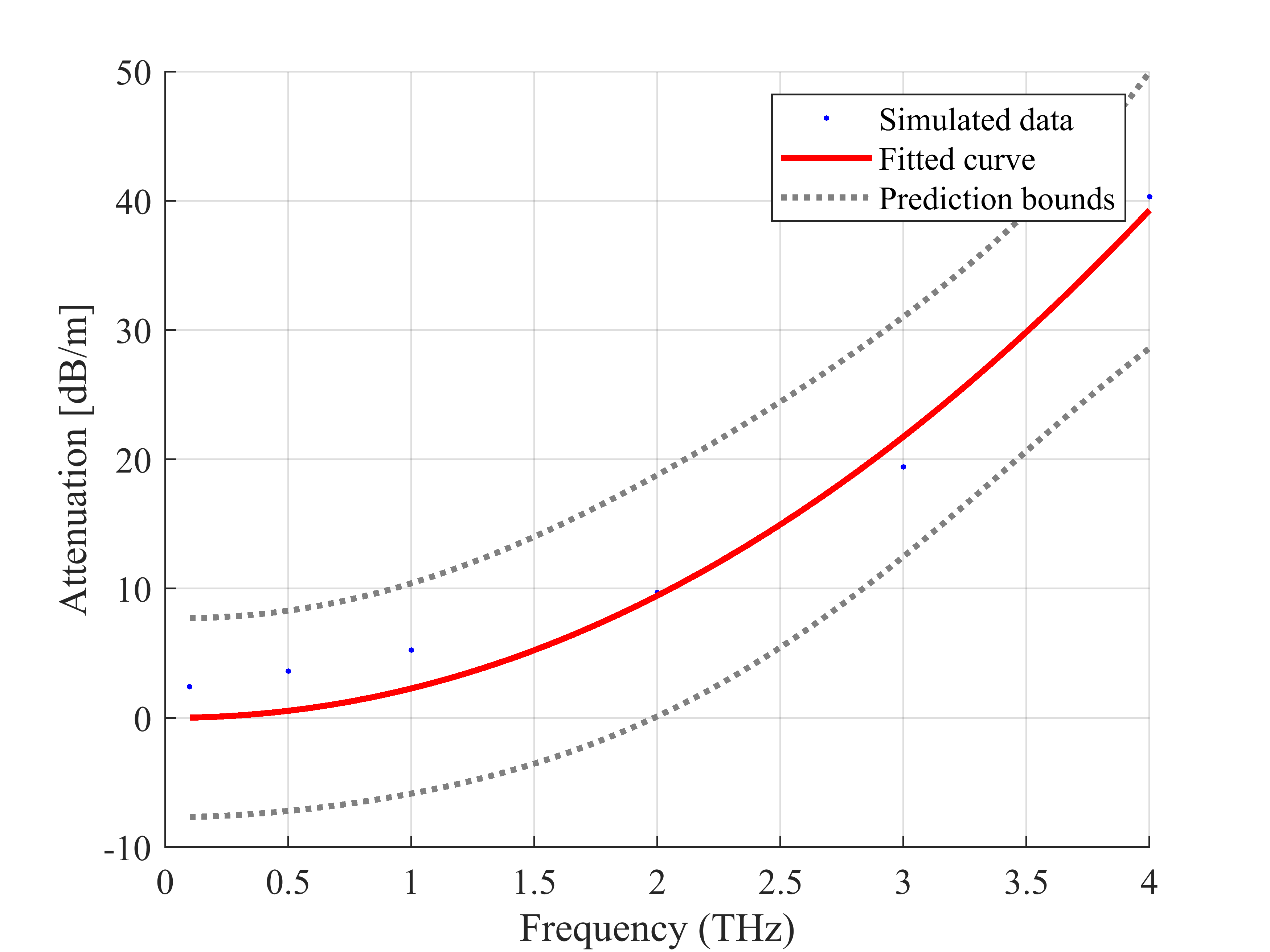}}
    \subfigure[Martian: Transmittance.]{\includegraphics[width=0.24\textwidth]{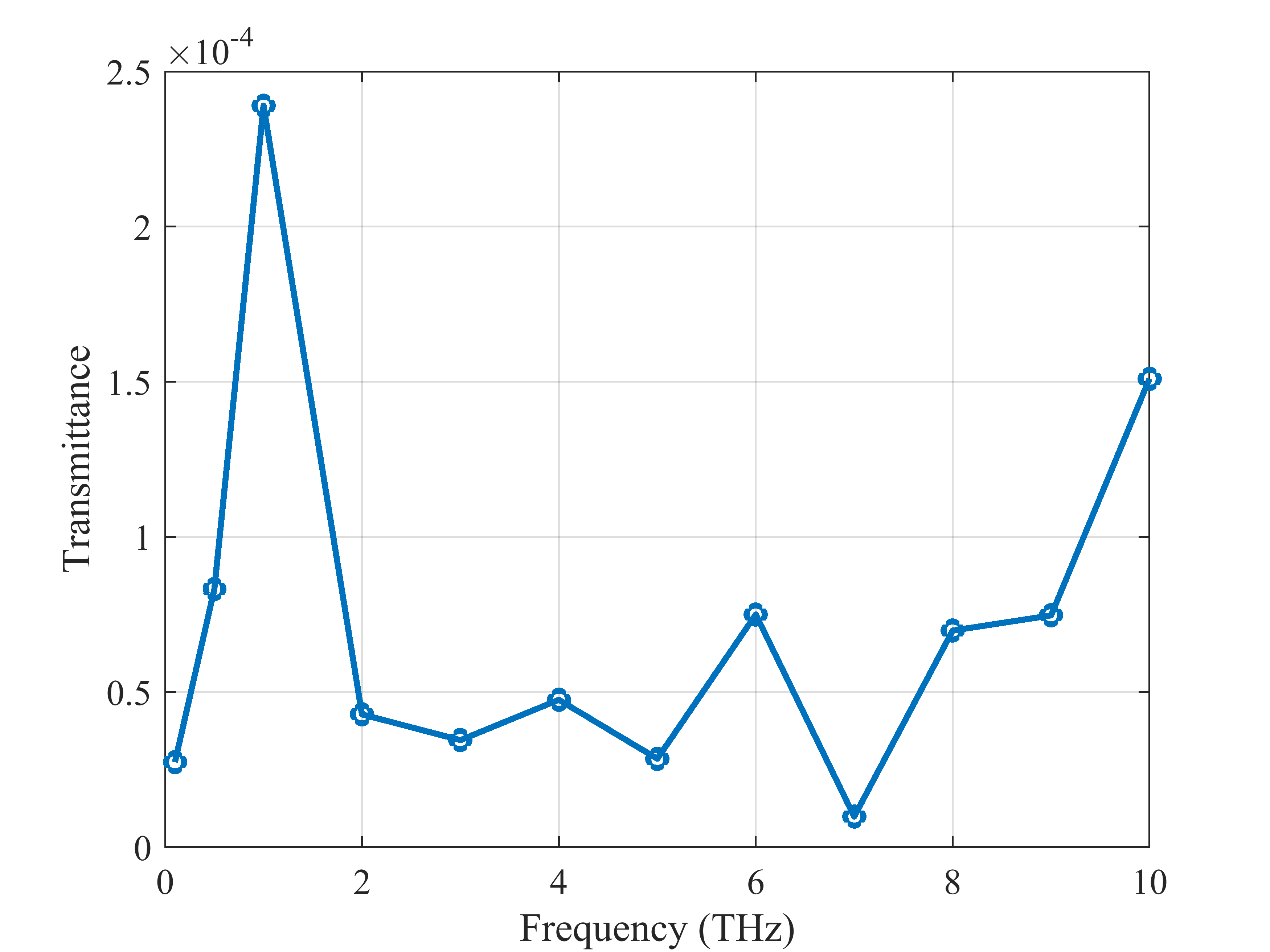}} 
    \subfigure[Martian: Attenuation ($dB/m$).]{\includegraphics[width=0.24\textwidth]{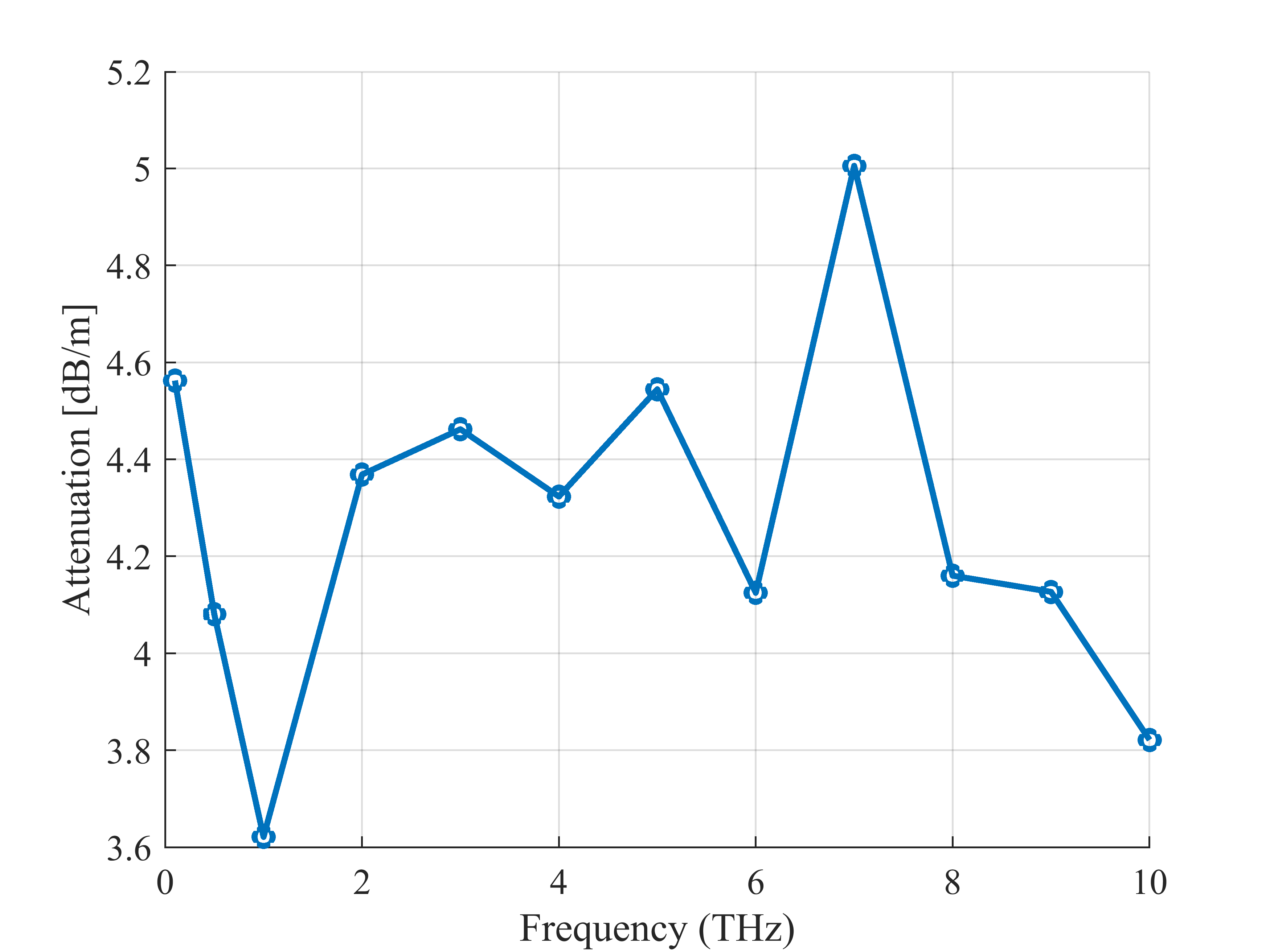}}
    \caption{Simulation for a THz beam by varying the frequency
from 0.1 to 10 THz while fixing dust particle number (100)
on the beam propagation path, MCP packets (10000), and the distance (10 $m$) between the transmitter and the receiver.}
    \label{fig:fre}
\end{figure} 

 \begin{figure*}[t]
    \subfigure[Particle number count.]{\includegraphics[width=0.32\textwidth]{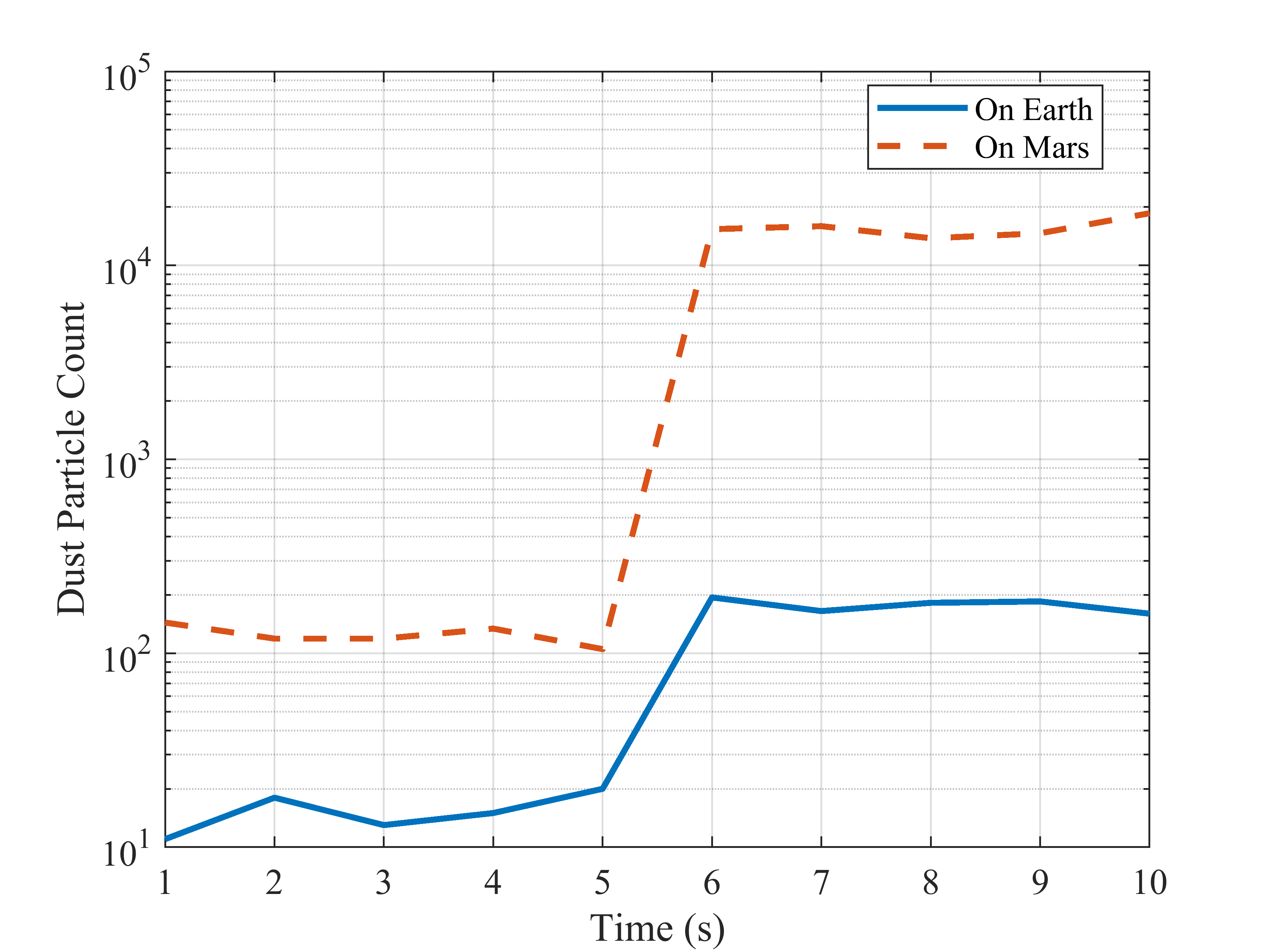}} 
    \subfigure[Transmittance.]{\includegraphics[width=0.32\textwidth]{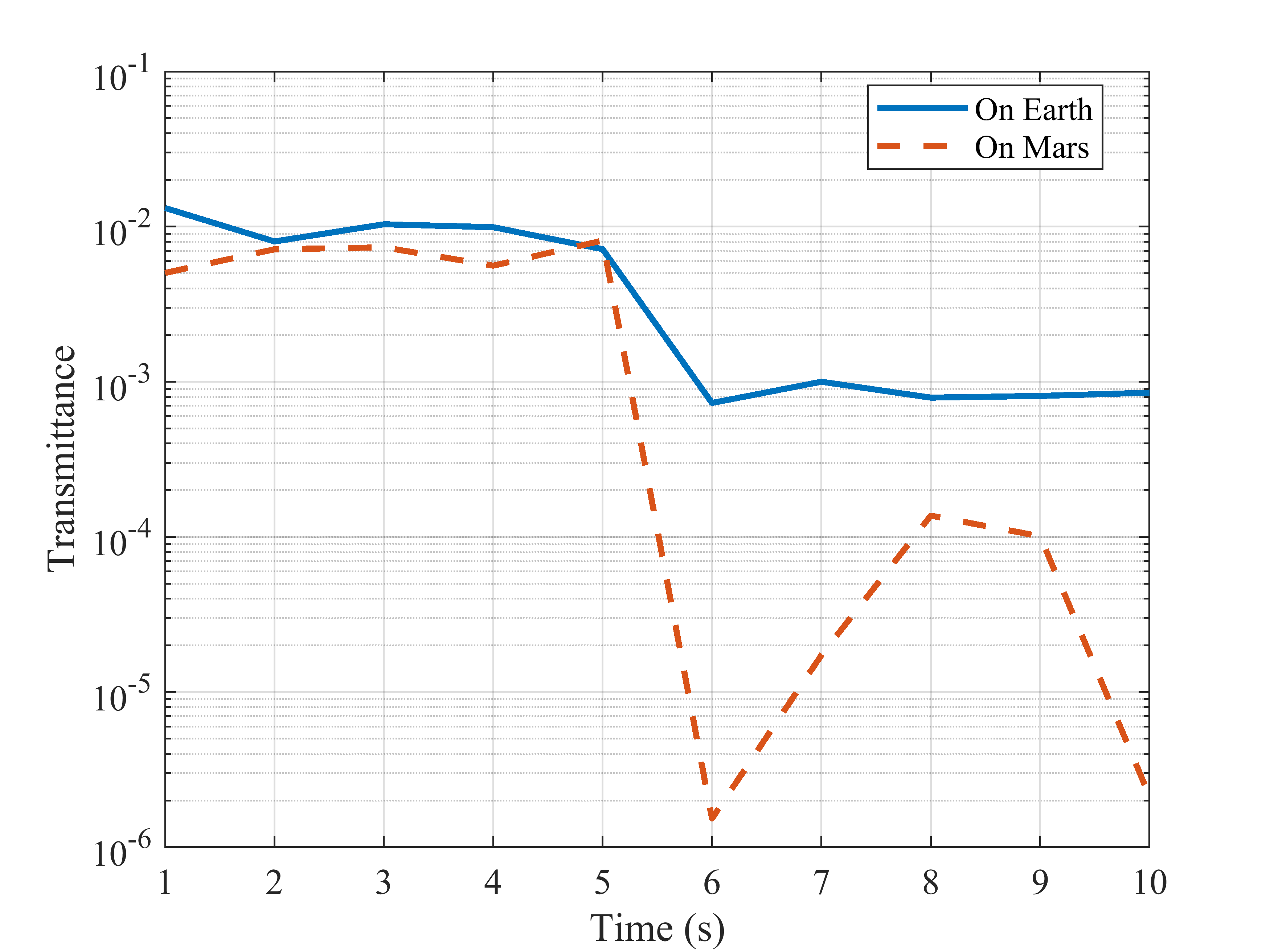}} 
    \subfigure[Attenuation ($dB/m$).]{\includegraphics[width=0.32\textwidth]{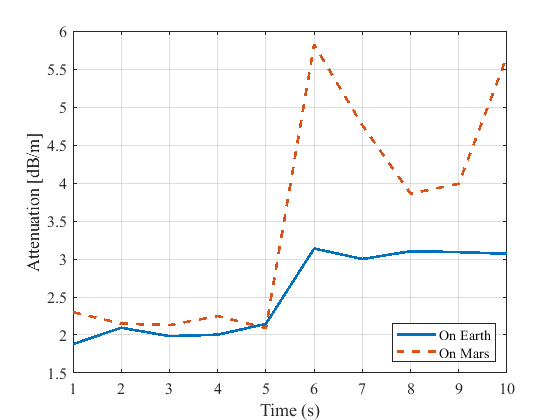}}
    \caption{Comparison of simulations for time-dependent turbulence (turmoil after 5 seconds) corresponding to the measurements of a) particle number count, b) the transmittance, and c) the attenuation ($dB/m$) considering Earth and Mars environments.}
    \label{fig:Time}
\end{figure*} 

Furthermore, we investigated the impact of frequency on the transmittance and attenuation measurements on Earth and Mars. As illustrated in Fig. \ref{fig:fre} (a), the transmittance measurements for Earth's environment decrease following an exponential function with the frequency increase from 0.1 to 4 THz. We were unable to calculate transmittance measurements following our simulation process beyond the 4 THz frequency limit since we are considering the dust particles with a radius of 1 to 150 microns for Earth environment, wavelengths can be comparably low or approximately equal to the dust particles' size after some frequencies threshold, creating more difficulties for data transmission. The corresponding attenuation for the transmittance measurements on Earth increases following a power function that can be fitted as $\text{Attenuation ($dB/m$)} = 2.277 D_{PN}^{2.054}$. On the other hand, the transmittance and attenuation measurements for the Mars environment do not show a particular increase or decrease trend. However, we noticed that the attenuation measurements vary around 4.2 $dB/m$ with the frequency increase from 0.1 to 10 THz.  

 Finally, we investigated the effect of time-dependent turbulence on the transmittance and the attenuation of the THz signal on Earth and Mars, which corresponds to the 0.24 THz and 1.64 THz frequencies, respectively and the distance between the transmitter and the receiver for 10 $m$. Here, we compare Earth and Mars simulation scenarios in which we assume that the dust particle number on the beam propagation path will suddenly increase due to the unpredictable behaviour of wind after 5 $s$. In the first 5 $s$, we assume that the dust particle number on the beam will vary between 10-20 in the Earth environment and 100-200 in the Mars environment in clear sky conditions. After five seconds, the dust particle number on the beam will increase between 100-200 on Earth and 10000-20000 on Mars due to the sudden wind turbulence in the communication area. As demonstrated in Fig. \ref{fig:Time} (a), dust particle number on the beam propagation path is low in the first 5 $s$ compared to the next 5 $s$ for both environments. Also, the dust particle number is higher for the Martian environment than the Earth environment in the considered time interval. When scrutinising the transmittance measurements (see Fig. \ref{fig:Time} (b)), we can notice that transmittance drops suddenly after 5 $s$ for both environments. However, average transmittance measurement values are approximately similar within the first 5 $s$. Also, the transmittance measurements on Mars after the turbulence are significantly low compared to Earth.  Corresponding to the transmittance measurements, attenuation measurements (see Fig. \ref{fig:Time} (c)) increases dramatically after 5 $s$ and are high on Mars, concluding that the turbulence effect on Mars should be investigated thoroughly. 

\begin{figure}[h]
    \centering
    \includegraphics[width=\linewidth]{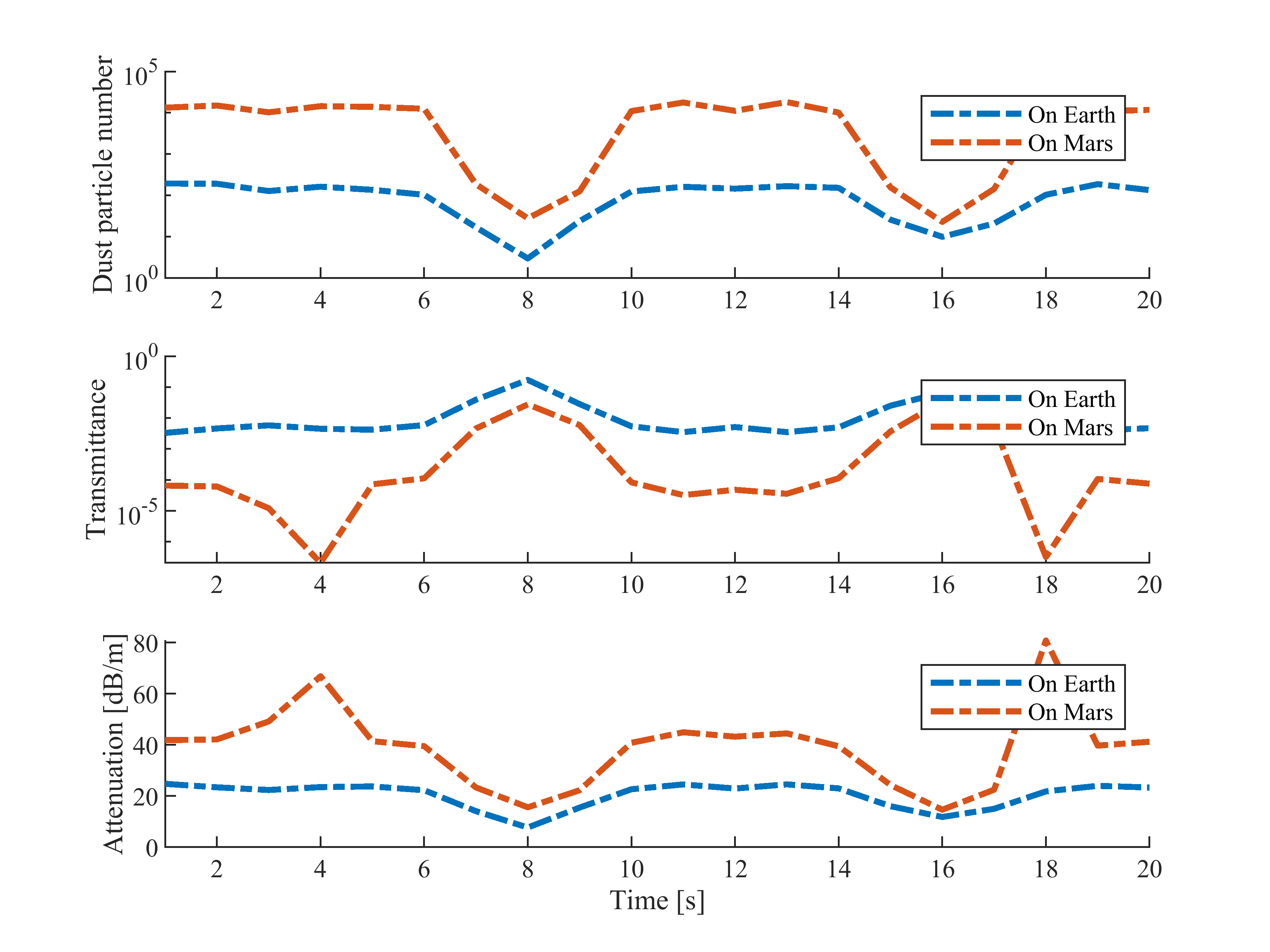}
    \caption{Measurements of Transmittance and Attenuation of 0.24 THz and 1.64 THz
links for Earth and Mars due to the sudden movement of dust particles on the beam propagation path for a fixed transmitter and receiver distance of 1 $m$.}
    \label{Fig10}
\end{figure}

\subsection{Channel Capacity simulation for Earth and Mars Under Dust storm}

\begin{figure}[h]
    \centering
    \includegraphics[width=\linewidth]{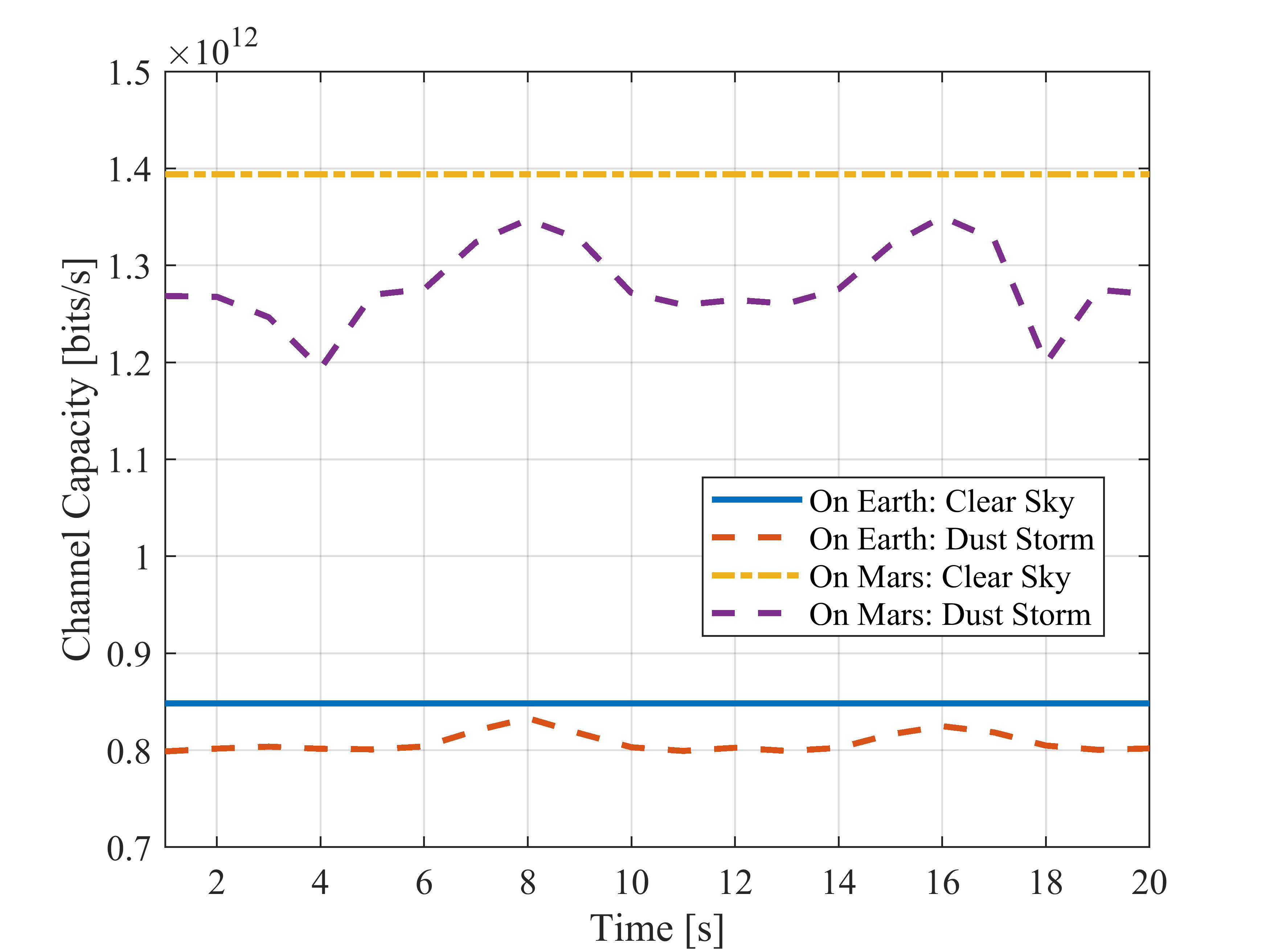}
    \caption{Channel capacity variation comparison for the scenario of a sudden drop in the number of dust particles on the beam due to the wind behaviour on Earth and Mars environments with considering spreading loss and molecular absorption loss.}
    \label{Fig11}
\end{figure}
This subsection investigates the channel capacity measurement of the THz links considering two scenarios. In the first scenario,  we assume that there are time windows when the number of dust particles on the THz beam propagation path drops, creating opportunities to communicate with high data rates for both Earth (0.24 THz) and Mars (1.64 THz) environments. Here we analyse the channel capacity for Earth and Mars environments considering spreading loss and Molecular absorption loss with the THz attenuation due to dust on the beam propagation path. Also, we investigate channel capacity variations in this scenario for different transmitter powers. In the second scenario, we investigate the channel capacity variation with the distance in clear sky and dust storm conditions.

In our first model, we assume that the dust density varies randomly for the Earth environment from 100 and 200 and Mars environment from 10000 to 20000 particles corresponding to a 1 $m$ distance between the transmitter and the receiver. Here we know that considering a 1 $m$ distance for simulation is unrealistic. However, in this scenario, we need to infer the effect of the sudden dust particle drop for the channel capacity measurements. Therefore, it is adequate to consider a 1 $m$ distance for the experiment. As we mentioned above, the significant variation of dust particles on Earth and Mars is in its effective radius. On Earth, the average effective radius of a dust particle varies between 1 and 150 microns \cite{Montecs}, and on Mars, it varies between 0.5 to 4 microns \cite{Dustproperties}. Therefore, to measure transmittance/attenuation considering approximately similar beam-blocking areas by dust particles on both Earth and Mars, we should consider 100 times more dust on Mars than on Earth, following the relationship between dust effective radius and area. Moreover, we sampled the dust particle number for each environment every second for 20 $s$. In addition, we assume there are two-time intervals (t=[7 $s$,9 $s$], t=[15 $s$,17 $s$]) when the dust particle number drops to less than 30 and 300 particles (see Fig. \ref{Fig10}) due to the unpredictable behaviour of wind. Such time intervals might represent occasions when the wind that causes dust particles to be suspended in the atmosphere falls away to near zero.

Furthermore, as we can see from Fig. \ref{Fig10}, the relative dust particle density decrease at the interval centred on t=16 $s$ is much higher than at the interval centred on t=8 $s$  for both environments. We noticed that corresponding to the low dust density, the transmittance measurement is higher at the interval 15-17 $s$ than at the 7-9 $s$ interval, and the attenuation shows the opposite variation to transmittance. In principle, those time intervals represent attractive time windows for communication in both environments because of lower attenuation due to the momentary absence of dust particles in the channel.

\begin{figure}[h]
    \centering
    \subfigure[On Earth]{\includegraphics[width=0.24\textwidth]{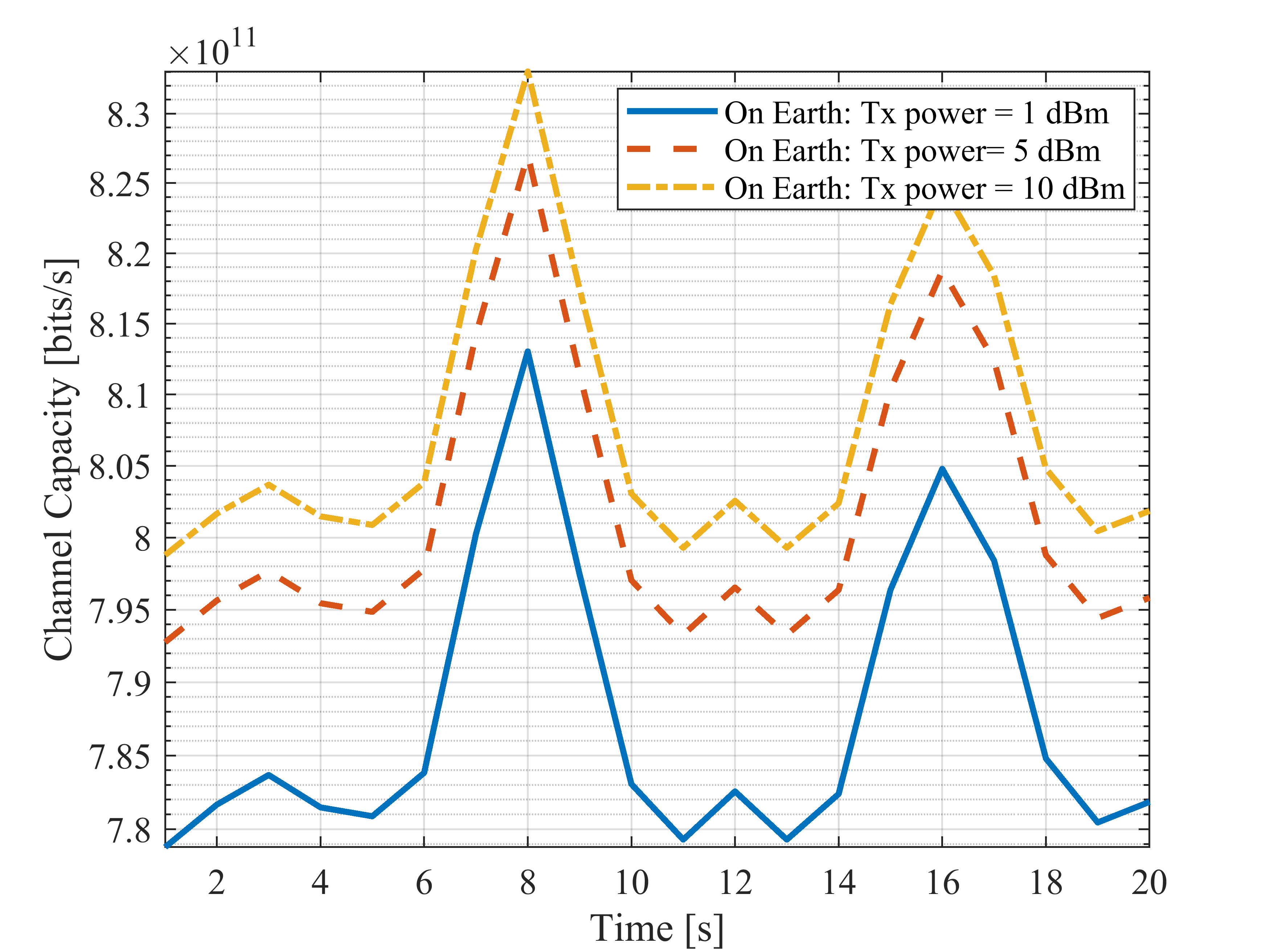}}
    \subfigure[On Mars]{\includegraphics[width=0.24\textwidth]{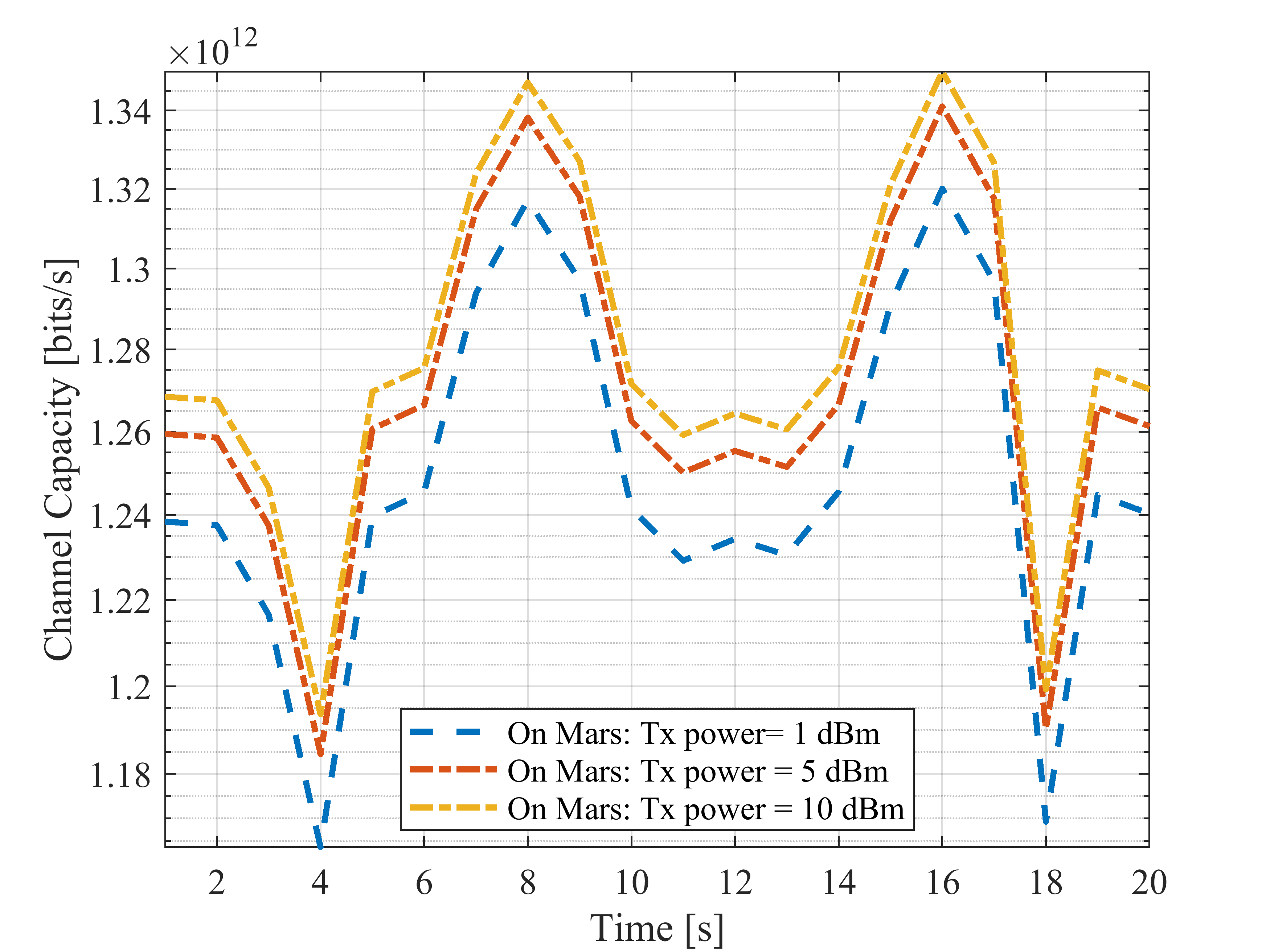}}
    \caption{Channel capacity measurement variation for the scenario of a sudden drop in the number of dust particles on the beam due to the wind behaviour for different Transmitted powers (1, 5, 10 $dBm$) of antenna with considering spreading and molecular absorption (a) on Earth and (b) on Mars.}
    \label{Fig12}
\end{figure}
Figure \ref{Fig11} illustrates the channel capacity variation in the clear sky and dust storm conditions on Earth and Mars for the same dusty scenario. When scrutinising Fig. \ref{Fig11}, we noticed that in clear sky conditions, channel capacity on Mars is approximately 1.39 $\times 10^{12}$ $bits/s$, and on Earth, its nearly 0.85 $\times 10^{12}$ $bits/s$. This shows more than 550 $GB/s$ difference between the channel capacity measurement in clear sky conditions on Mars and Earth due to much higher molecular absorption on Earth. Also, it is noticeable that dust appears to have a more significant relative effect on channel capacity on Mars than on Earth in a dust storm situation due to the high number of tiny dust particles on the beam propagation path. Moreover, channel capacity measurements in dust storm situation on Earth is lower than in clear sky condition, but the difference is minimal. However, the difference on Mars is relatively enormous, and it is approximately two orders of magnitude less than clear sky conditions. On the other hand, channel capacity measurements in dust storm condition on Mars is higher by approximately five orders of magnitude than in dust storm condition on Earth. In addition, the free space path loss is constant in this scenario because it only depends on the carrier frequency and the distance between the transmitter and the receiver, which are both constant in this simulation process.

In Fig. \ref{Fig12}, we investigated the channel capacity measurement variations for different transmitter power in discrete time windows considering the free space path loss and the molecular absorption loss effect on the channel. This figure shows that the channel capacity increases as antenna transmitter power increases for both Earth and Mars environments by approximately 5 to 20 $GB/s$. Also, we noticed that we could have reliable communication links with high channel capacities when communicating in the time windows with low dust densities on Mars. Moreover, the channel capacity measurements show a significant variation on Earth compared to Mars for all transmitted powers. Therefore, these measurements imply that we can reach high channel capacities on Mars than on Earth using lower transmitter power antennas.

\begin{figure}[!htb]
    \centering
    \includegraphics[width=\linewidth]{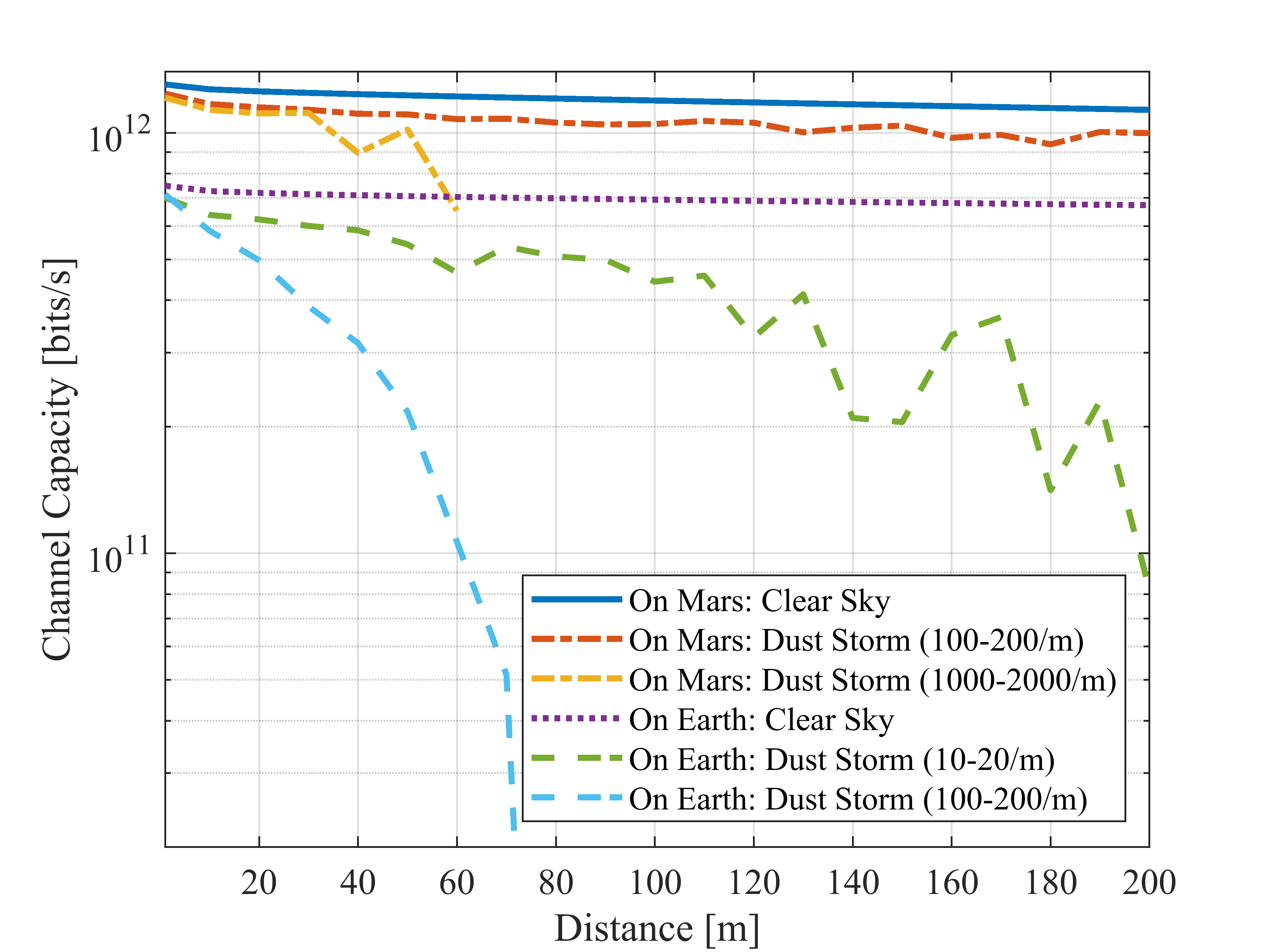}
    \caption{Channel capacity measurements by varying the distance between the transmitter and the receiver with and without dust storms situations on Mars and Earth.}
    \label{Fig13}
\end{figure}

In our second and final simulation scenario, we investigated the channel capacity for various distances between the transmitter and the receiver for both Earth and Mars environments, comparing clear sky and dust storm conditions with different dust densities per meter (See Fig. \ref{Fig13}). The transmitter power was taken as 10 $dBm$ in this simulation. Here, we allowed the dust particle count density to vary as a factor of distance to simulate more realistic dust storm conditions. For the Earth's environment, we have considered dust storms that result in 10-20 (very low) and 100-200 (very high) dust particles per meter dust densities on the THz beam propagation path. Similar conditions for the mars environment were considered by comparing the dust particle sizes with Earth. Thus, we assumed that dust storms on Mars would carry 100-200 and 1000-2000 dust particles per meter to the beam propagation path. However, we should have considered 10000-20000 dust particles per meter on Mars. Nevertheless, due to computational difficulties with high dust densities with increasing the distance between the transmitter and the receiver, we are considering above mentioned numbers for channel capacity measurements. Moreover, we have taken account of spreading loss and molecular absorption loss when calculating the channel capacities for each distance.

As shown in Fig. \ref{Fig13}, the channel capacity decreases gradually for clear sky conditions for both environments showing high channel capacities in the considering distance range. However, the decrement is high on Earth due to high molecular absorption. Moreover, the channel capacity measurements decrease dramatically on Earth with the distance for 100-200/$m$ dust storm conditions, showing communication blackout beyond 70 $m$ distance. Also, at 10-20/$m$ dust storm, channel capacity measurements decrease slowly, showing that this dust particle number on the beam propagation path is not a massive issue for achieving high channel capacities. However, Earth's channel capacity is significantly dropping when compared with the channel capacity decrement on Mars for the 100-200/$m$ dust particle density. Again, 100-200/$m$ dust particles on the beam propagation path on Mars do not significantly affect the channel capacity measurement. Therefore, we can neglect the THz link budget degradation due to the small amount of dust on Mars. In addition, when investigating the high dust particle number density effect on the channel capacity on Mars (1000-2000/$m$), the channel capacity measurements are computationally difficult when the distance is greater than 60 $m$. However, we can notice that the channel capacity measurements for Mars are decreasing rapidly with the increase in the distance. Also, we can see that the channel capacities are equal at a distance of approximately 60 $m$ for clear sky conditions on Earth and the high dust density scenario on Mars.

\section{Conclusion}
\label{conclusion}
High-speed, reliable communication between devices on Earth and Mars is needed to fulfil future communication requirements. In this study, we investigated the impact of atmospheric dust and dust storms for communication using THz links, utilising a modified Monte Carlo simulation algorithm. The calculated transmittance and attenuation measurements are based on Mie and Rayleigh approximations depending on the dust particle sizes and carrier frequency utilised for communication on the two planets. Moreover, we presented a channel capacity model and analysed it for two different time-dependent and distance-dependent scenarios. The Monte-Carlo simulation results show that attenuation measurements decrease for both Earth and Mars environments when the MCP packets and visibility increase. In addition, for both environments, the attenuation increases with higher dust particle number on the beam propagation path and distance between the transmitter and the receiver. We noticed the exact attenuation behaviour with the increased frequency for the Earth's environment. However, the attenuation measurements vary around a constant value for the Mars environment. When scrutinising the channel capacity measurements from the time-dependent scenario, we can conclude that the time windows showing sudden dust particle density drops create the best communication opportunities with high data rates. Also, we noticed that the channel capacity measurements dramatically drop with the increase in distance between the transmitter and the receiver in severe dust storm situations on both Earth (100-200/$m$) and Mars (1000-2000/$m$) environments, even if we use low molecular absorption frequencies and high transmitter power antennas. However, the impact from the local dust storm is negligible on Mars (100-200/$m$) but should be further investigated on Earth (10-20/$m$). 

\section*{Acknowledgment}

This publication came from research conducted with the financial support of Science Foundation Ireland (SFI) and the Department of Agriculture, Food and Marine on behalf of the Government of Ireland (Grant Number [16/RC/3835] - VistaMilk), the support of YL Verkot, Finland, and US National Science Foundation (NSF) ECCS-2030272 grant.

\ifCLASSOPTIONcaptionsoff
  \newpage
\fi

\bibliographystyle{IEEEtran}
\bibliography{References}

\begin{IEEEbiography}[{\includegraphics[width=1in,height=1.25in,clip,keepaspectratio]{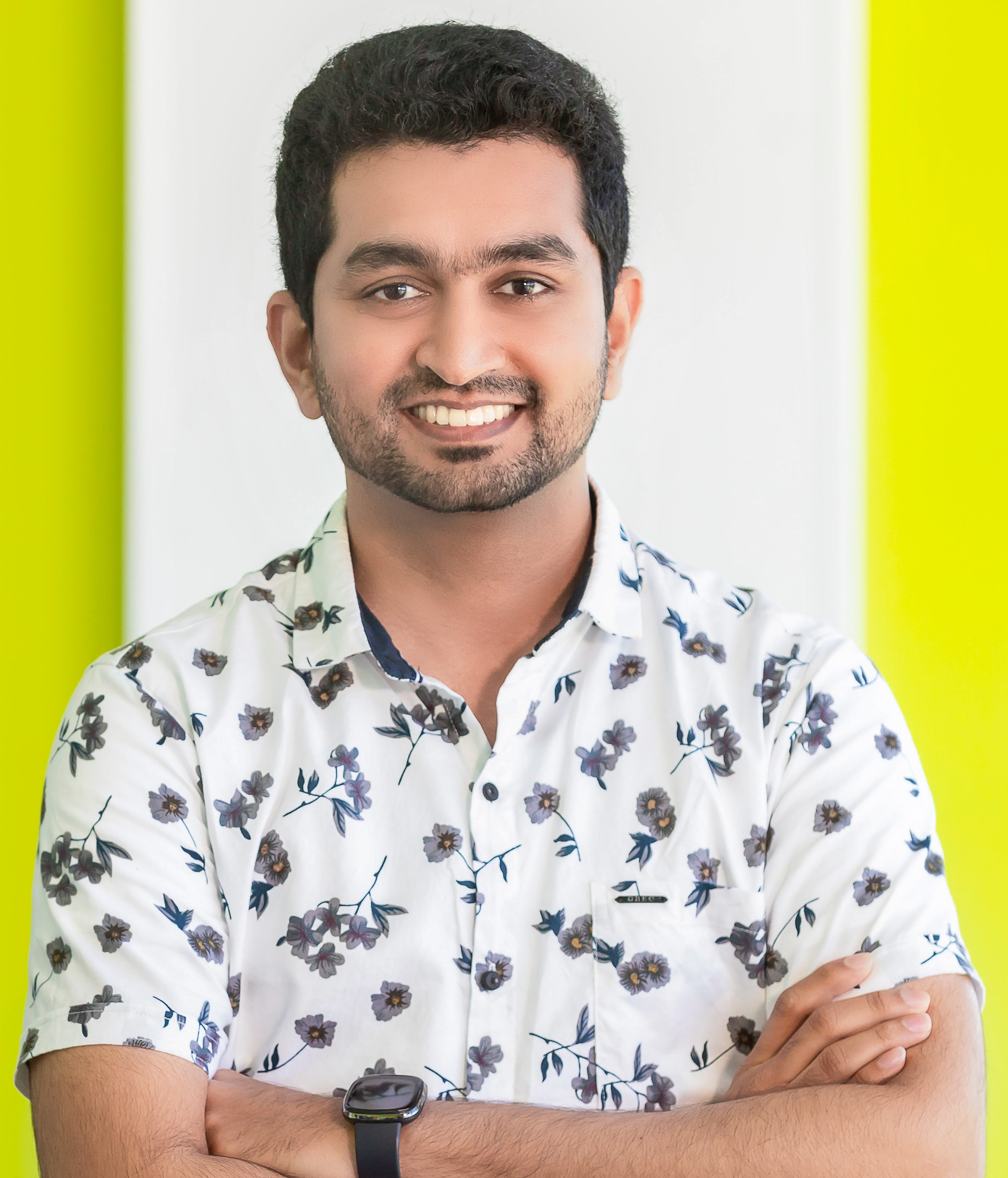}}]{LASANTHA THAKSHILA WEDAGE} [S'22] (thakshila.wedage@waltoninstitute.ie)
received his B.S. degree in Mathematics from University of Ruhuna, Sri Lanka, in 2016. 
He is currently pursuing a PhD degree with the Department of Computing and Mathematics, Walton Institute, South East technological University, Ireland. His current research interests lie in Mathematical modelling and 5G/6G Wireless communication and sensing.
\end{IEEEbiography}

\begin{IEEEbiography}[{\includegraphics[width=1in,height=1.25in,clip,keepaspectratio]{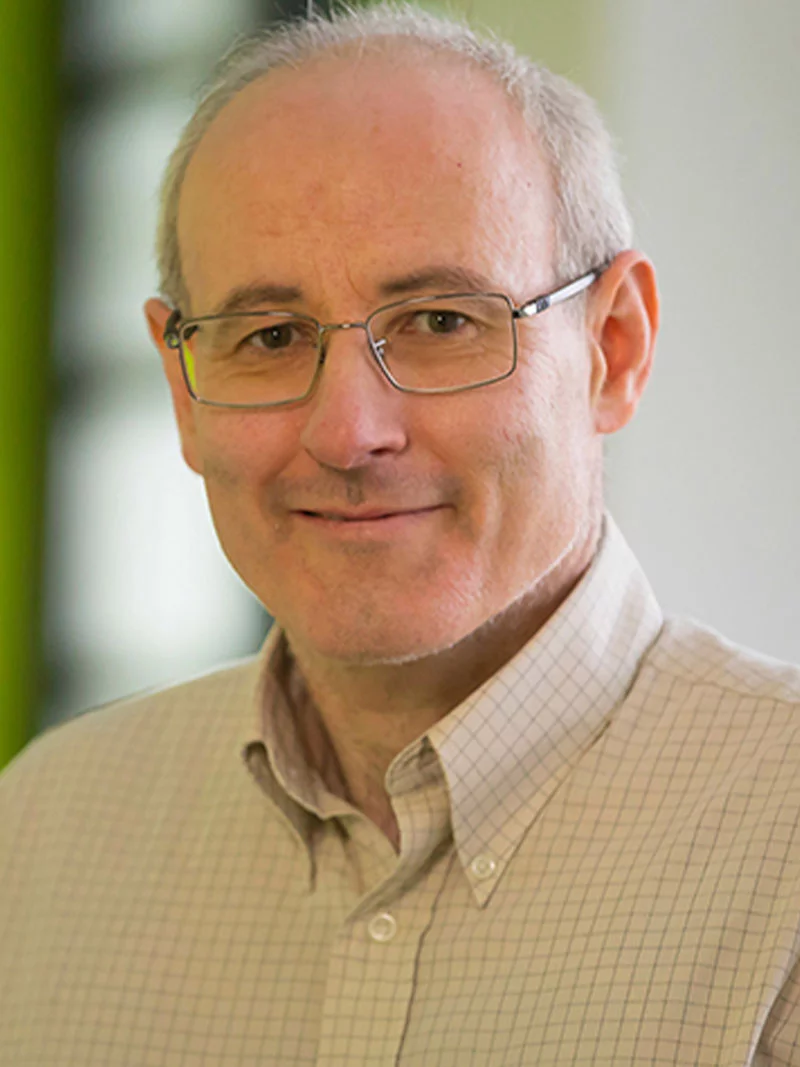}}]{BERNARD BUTLER}[SM'22] (bernard.butler@setu.ie) received his PhD from South East Technological University, Ireland. He was a Senior Research Scientist in the U.K.’s National Physical Laboratory, focusing on mathematics of measurement and sensing. He is a Lecturer in SETU and is CONNECT Funded Investigator and VistaMilk Academic Collaborator with the Walton Institute, SETU. Research interests include machine learning, wireless comms and edge networking.
\end{IEEEbiography}

\begin{IEEEbiography}[{\includegraphics[width=1in,height=1.25in,clip,keepaspectratio]{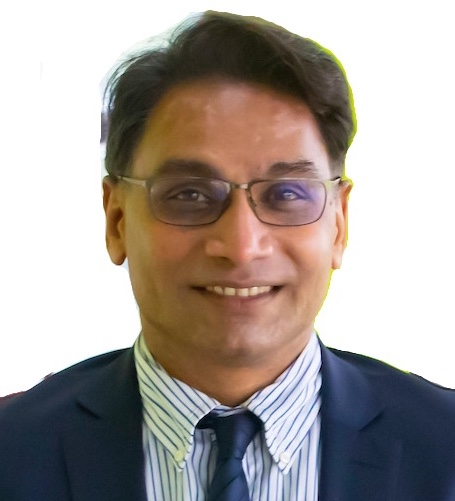}}]{SASITHARAN BALASUBRAMANIAM} [SM'14] (sasi@unl.edu) received his Bachelors in Engineering and PhD degree from the University of Queensland, Australia in 1998 and 2005, respectively, and Masters of Engineering Science from Queensland University of Technology in 1999. He was a past recipient of the Science Foundation Ireland Starter Investigator Research Grant. He was also a past recipient of the Academy of Finland Research Fellow at Tampere University, Finland. He was previously the Director of Research at the Walton Institute, South East Technological University, Ireland. He is currently an Associate Professor at the School of Computing, University of Nebraska-Lincoln. He is currently the Editor-in-Chief of IEEE Transactions on Molecular, Biological and Multi-scale Communications as well as an Associate Editor for IEEE Transactions on Mobile Computing. He was an IEEE Distinguished Lecturer for the IEEE Nanotechnology Council in 2018. His research interests lie in molecular and nano communications, Internet of Bio-Nano Things, as well as 5G/6G networks.
\end{IEEEbiography}

\begin{IEEEbiography}[{\includegraphics[width=1in,height=1.25in,clip,keepaspectratio]{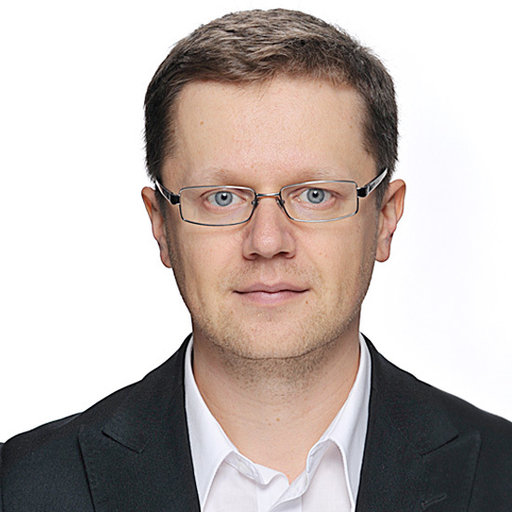}}]{YEVGENI KOUCHERYAVY}[SM'08] (yevgeni.koucheryavy@yl-verkot.com)
received the Ph.D. degree from the Tampere University of Technology, Finland, in 2004. He is currently a Full Professor with the Unit of Electrical Engineering, Tampere University, Finland. He has authored numerous publications in the field of advanced wired and wireless networking and communications. His current research interests include various aspects in heterogeneous wireless communication networks and systems, the Internet of Things and its standardization, and nanocommunications.
\end{IEEEbiography}

\begin{IEEEbiography}[{\includegraphics[width=1in,height=1.25in,clip,keepaspectratio]{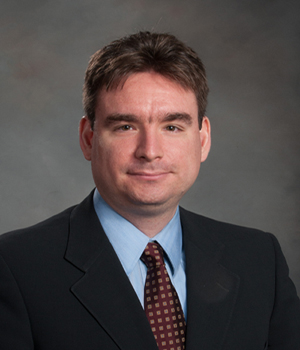}}]{Mehmet Can Vuran}[M'07] (mcv@unl.edu) was born in Istanbul, Turkey. He received his B.Sc. degree in Electrical and Electronics Engineering from Bilkent University, Ankara, Turkey, in 2002. He received his M.S. and Ph.D. degrees in Electrical and Computer Engineering from Georgia Institute of Technology, Atlanta, GA., in 2004 and 2007, respectively. Currently, he is the Dale M. Jensen Chair Professor in Computing at the School of Computing at the University of Nebraska-Lincoln. Dr. Vuran has been recognized as a Highly Cited Researcher three years in a row by Thomson Reuters "in recognition of ranking among the top 1\% of researchers for most cited documents in Computer Science". Dr. Vuran was awarded an NSF CAREER award for the project ``Bringing Wireless Sensor Networks Underground''. He is a Daugherty Water of Food Institute Fellow and a National Strategic Research Institute Fellow. He serves on the editorial boards of IEEE Transactions on Wireless Communications, IEEE Transactions on Mobile Computing, and IEEE Transactions on Network Science and Engineering. His research interests are in 6G networks, the Internet of Things (IoT), agricultural wireless networks, wireless underground communications, and vehicular communications.
\end{IEEEbiography}

\end{document}

%% file: intro.tex
\section{Introduction}
\label{intro}

\IEEEPARstart{S}{ixth} generation (6G) wireless networks aim to push the frequency spectrum into the Terahertz (THz) band to fulfill rising capacity demands and requirements, given the opportunity for higher bandwidths \cite{shafie2022terahertz, AkyildizJosep2022,Han2022}. The 0.1  to 10 THz frequency range has the potential to (1) realize high bandwidth transmissions that can allow hundreds of GB/s data rates for communication \cite{2022terahertz_1,Hang2022,shafie2021spectrum}, and (2) provide new opportunities to create miniature THz-enabled antennas due to the small wavelengths (30 $\mu$m – 3 mm), enabling us to design arrays with a large number of antenna units \cite{jornet2013graphene,SasiBernard2020digital,sasi2022}. Numerous studies have shown that specific THz frequencies suffer high molecular absorption due to atmospheric gases (e.g., water vapor and oxygen). However, given the wavelength and high energy photons of THz signals, other particles can also significantly impact the link budget, which can result in scattering and absorption of signal power. 

Recent studies have shown that solid particles such as dust, sand and ice affect THz signals \cite{snow}, in addition to molecular absorption from atmospheric gases \cite{JosepTeraNets2014}. However, past studies have paid little attention to signal attenuation caused by solid particles such as dust and sand. Therefore, further investigation is required to determine how dynamic environments composed of solid particles, such as dust storms, affect THz links. This requires further investigation, especially as we expand connectivity in rural areas and other planets (e.g., Mars) to interplanetary scale. In the case of Mars, the recent vision of colonizing the planet will require high-bandwidth connectivity to maximize chances for human survival.   

A dust storm is a physical layer of dust and debris blown into the atmosphere by winds with horizontal and vertical velocity components. On Earth, the wall of dust can be miles wide and several thousand feet high. Dust storms are more frequently found in arid regions such as the Middle East \cite{MiddleEast}, North China \cite{China_2022}, and North Africa \cite{america_2020} at specific periods of the year. In more densely populated areas, human activity creates dust when burning fossil fuels for heating, cooking, or transport. Industrial and construction processes also create dust. This study compares the effects of solid dust particles on (sub)THz signals on both Earth and Mars, taking account of varying environmental conditions. Considering the differences in atmospheric conditions on Earth and Mars, with or without dust, suggests the use of different frequencies to enable relatively long-distance wireless communication on both planets. Dust storms are one of the most remarkable features on Mars. Even though wind speed on Mars is not significantly higher than on Earth, the extremely dry, dusty surface yields more dust storms.

\begin{figure*}[t]
    \centering
    \includegraphics[width=\linewidth]{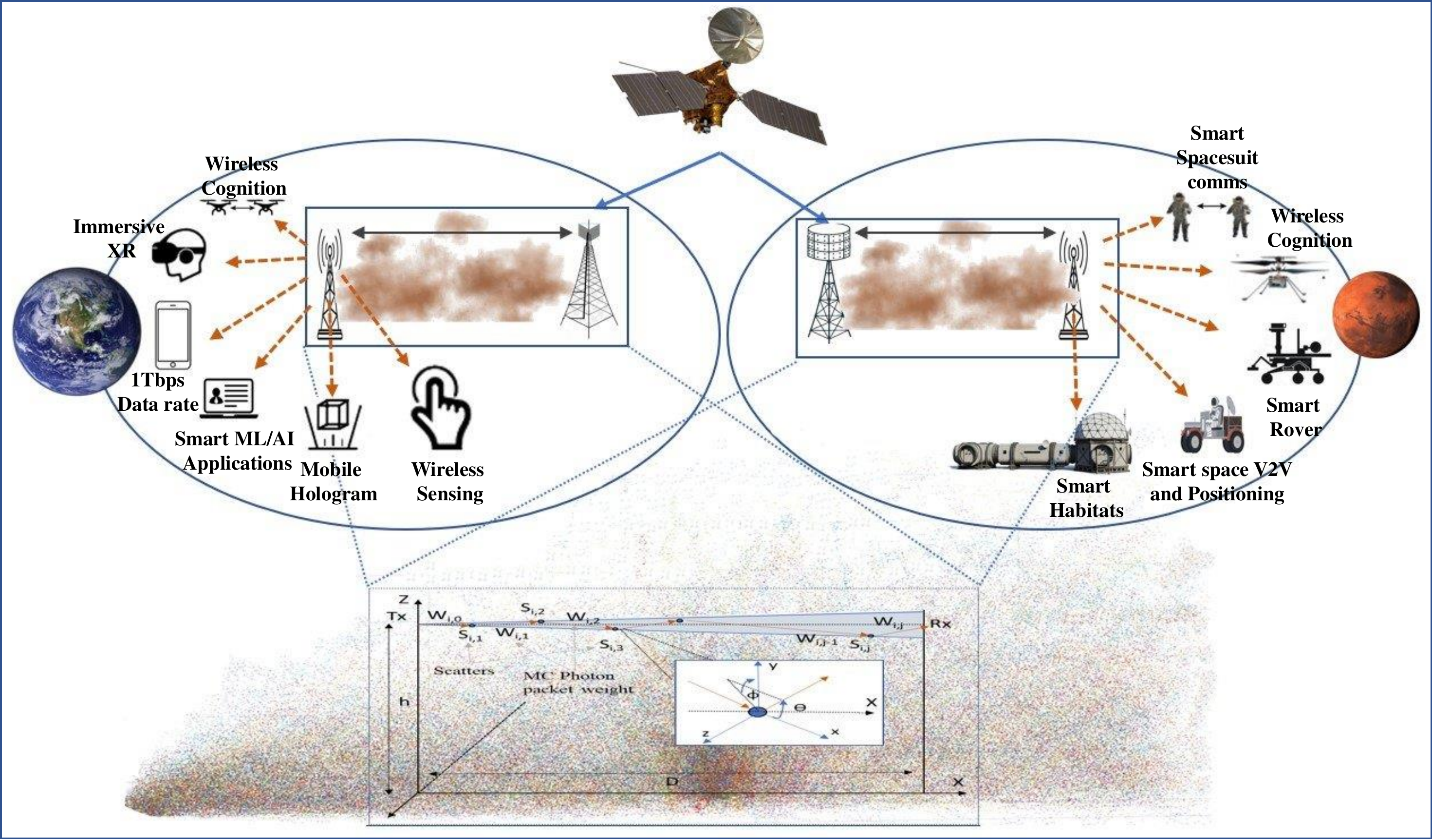}
    \caption{THz wireless communication applications and links through Earth and Mars atmospheric and environmental conditions.}
    \label{fig:Overview}
\end{figure*}

Figure~\ref{fig:Overview} provides an overview of \emph{selected} wireless communications applications on Earth, and \emph{proposed} wireless communication applications on Mars. While the applications differ, they are both affected by wireless channel losses, including those caused by dust particles that can scatter the EM waves used for communication. The rest of this paper considers the similarities and differences in the channel conditions, and includes models and simulations based on simulated dust storms that result in beam scattering, as shown at the bottom of Fig.~\ref{fig:Overview}.

In a dusty environment, the dust particle density is higher than usual, and the effects of multiple scattering of EM waves due to dust particles are non-negligible. Recent studies have not considered this significant effect on the attenuation of EM waves \cite{single2018scattering}, \cite{ep_value}. The lack of consideration of multiple scattering effects can result in significant gaps between theoretical and experimental results. This paper considers  multiple scattering of EM waves due to dust particles along the beam propagation path. To this end, we model the EM wave as a $\textbf{\text{\textit{photon packet}}}$ instead of a shower of photons. It is inaccurate to consider the EM wave as a shower of photons characterized by the position of a photon and its trajectory \cite{mishchenko2009gustav}. A photon packet models a portion of the \emph{energy weight} of the EM wave rather than single photons (which have quantum behavior). Therefore, we can consider an EM wave as a collection of energy packets and model multiple scattering effects utilizing the Monte Carlo algorithm, to infer the radiative transfer equation. The THz link scattering loss measurement in this study is inspired by \cite{Montecs}, where the scattering loss due to charged dust particles is calculated by considering the energy of the transmitting signal as Monte Carlo Photon Packets. Vertical THz attenuation is determined in \cite{Montecs}, but this study considers horizontal point-to-point communication for both Earth and Mars in dusty atmospheric scenarios. 

    \begin{table}[t]
        \centering
        \caption{Atmospheric gas composition comparison between Earth and Mars \cite{ep_value}; ppm is a concentration of parts per million.}
        \begin{tabular}{|l|l|l|}
        \hline
            Gas & Composition on Earth &Composition on Mars\\
        \hline
          \ce{N2}   & 78.084\%      &2.7\% \\
          \ce{O2}   &  20.946\%     & 0.13\%\\
          \ce{Ar}   &  0.93\%       & 1.6\%\\
          \ce{H2O}  &  1-3\%       & 100-400$ppm$\\
          \ce{CO2}  &  0.003\%     & 95.32\%\\
          \ce{CH4}  &  1.5$ppm$   & -\\
          \ce{SO2} &  1$ppm$     & -\\
          \ce{O3}   &  0.05$ppm$   & 0.1$ppm$\\
          \ce{N2O}  & 0.02$ppm$     & -\\
          \ce{CO}  &  0.01$ppm$  & 0.08\% \\
          \ce{NH3}  & 0.01$ppm$     & -\\
          \ce{NO}   & -             & 100$ppm$\\
        \hline
        \end{tabular}
        \label{tab:AtmosphereComparison}
    \end{table}

The contributions of this paper are:
\begin{enumerate}
    \item A 3-D geometric scattering model for multiple photon-dust particle interactions is presented, using both Mie and Rayleigh approximations, to estimate the probability that a photon packet arrives at the receiver.
    \item The model is used in simulation to estimate the overall channel capacity considering THz and sub-THz link budget degradation due to the combination of scattering by dust particles, molecular absorption loss due to the atmosphere, and free-space spreading loss.
    \item Different communication channel conditions (on Earth and Mars) and their effect on channel capacity, including power loss caused by multiple scattering by dust, are compared and analysed.
\end{enumerate}

The rest of this paper is organized as follows: 
Section~\ref{background} describes dust conditions and how they affect EM propagation and contrasts the conditions that prevail on Earth and Mars; Section \ref{3Dstorm} describes how 3D dust storm simulation is affected by the number of dust particles on the EM wave propagation path. Then Section \ref{MCPsim} explains the Monte-Carlo simulation process for calculating the transmittance/attenuation when photon packets are scattered by multiple dust particles. Section \ref{channelcap} presents estimates of transmittance/attenuation obtained by Monte-Carlo simulation, in various parameter settings. Section \ref{RnD} presents a channel capacity model that combines the effect of spreading, molecular absorption and multiple scattering by dust, with simulated results. Finally, Section \ref{conclusion} presents our conclusions.